\documentclass[sigconf]{acmart}
\usepackage[english]{babel}
\usepackage{blindtext}
\AtBeginDocument{%
  }
\renewcommand\footnotetextcopyrightpermission[1]{} % removes footnote with conference info
\setcopyright{none}
\copyrightyear{2026}
\acmYear{2026}
\acmDOI{XXXXXXX.XXXXXXX}
\acmConference[Preprint]{Make sure to enter the correct
  conference title from your rights confirmation email}{\xspace}{\xspace}
\acmISBN{978-1-4503-XXXX-X/2018/06}
\usepackage{natbib} 
\usepackage{amsmath}
\usepackage{xspace}
\usepackage{tikz}
\usepackage{graphicx}
\usepackage{pgf-pie}
\usepackage{pgfplots}
\usepackage{pifont}
\usepackage{enumerate}
\usepackage{enumitem}
\usepackage{booktabs}
\usepackage{tabularx}
\usepackage{caption}
\usepackage[dvipsnames]{xcolor}
\usepackage[export]{adjustbox}
\usepackage{url}
\usetikzlibrary{positioning, arrows.meta, shapes.geometric, fit, calc}
\newcommand{\fakepar}[1]{\par\addvspace{0.5em} \noindent \textbf{#1.}\xspace}
\newcommand{\geminiicon}{\includegraphics[height=1.8ex]{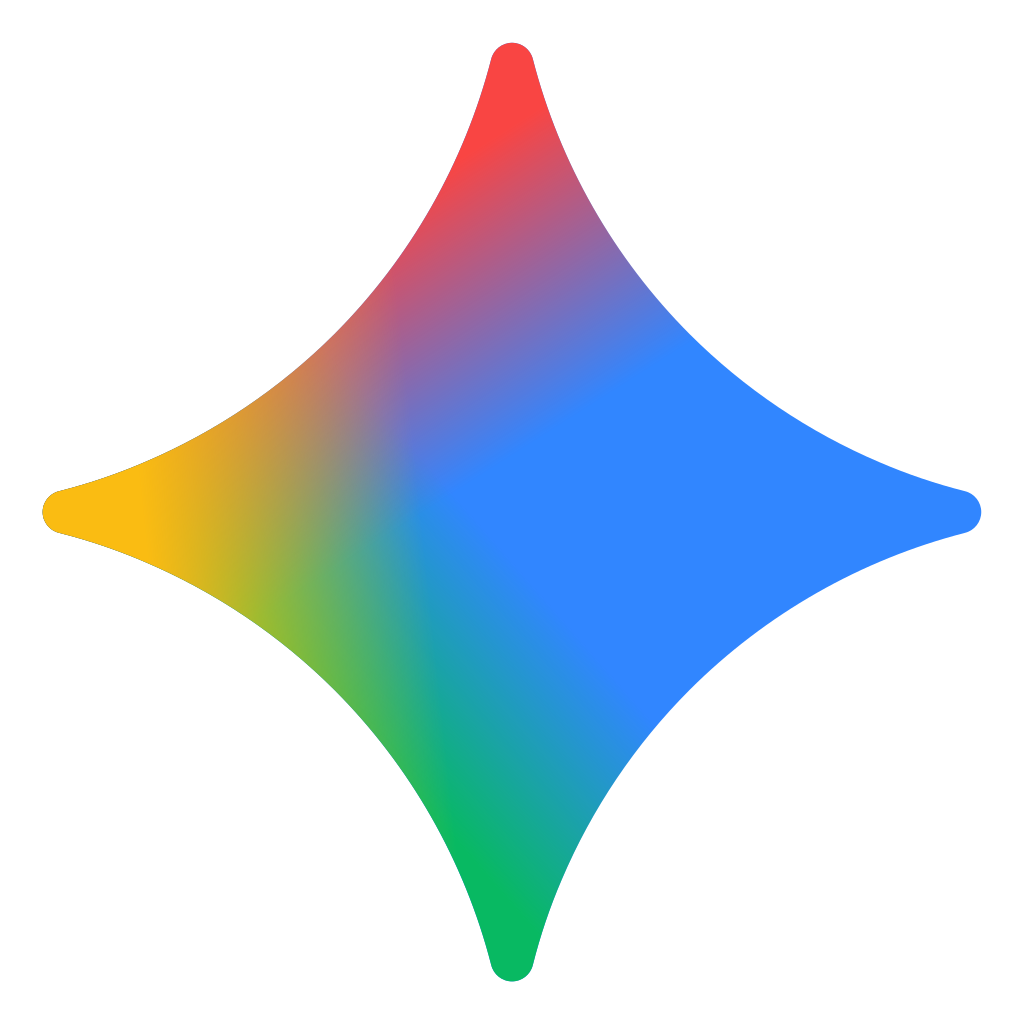}}
\newcommand{\gpticon}{\includegraphics[height=1.6ex]{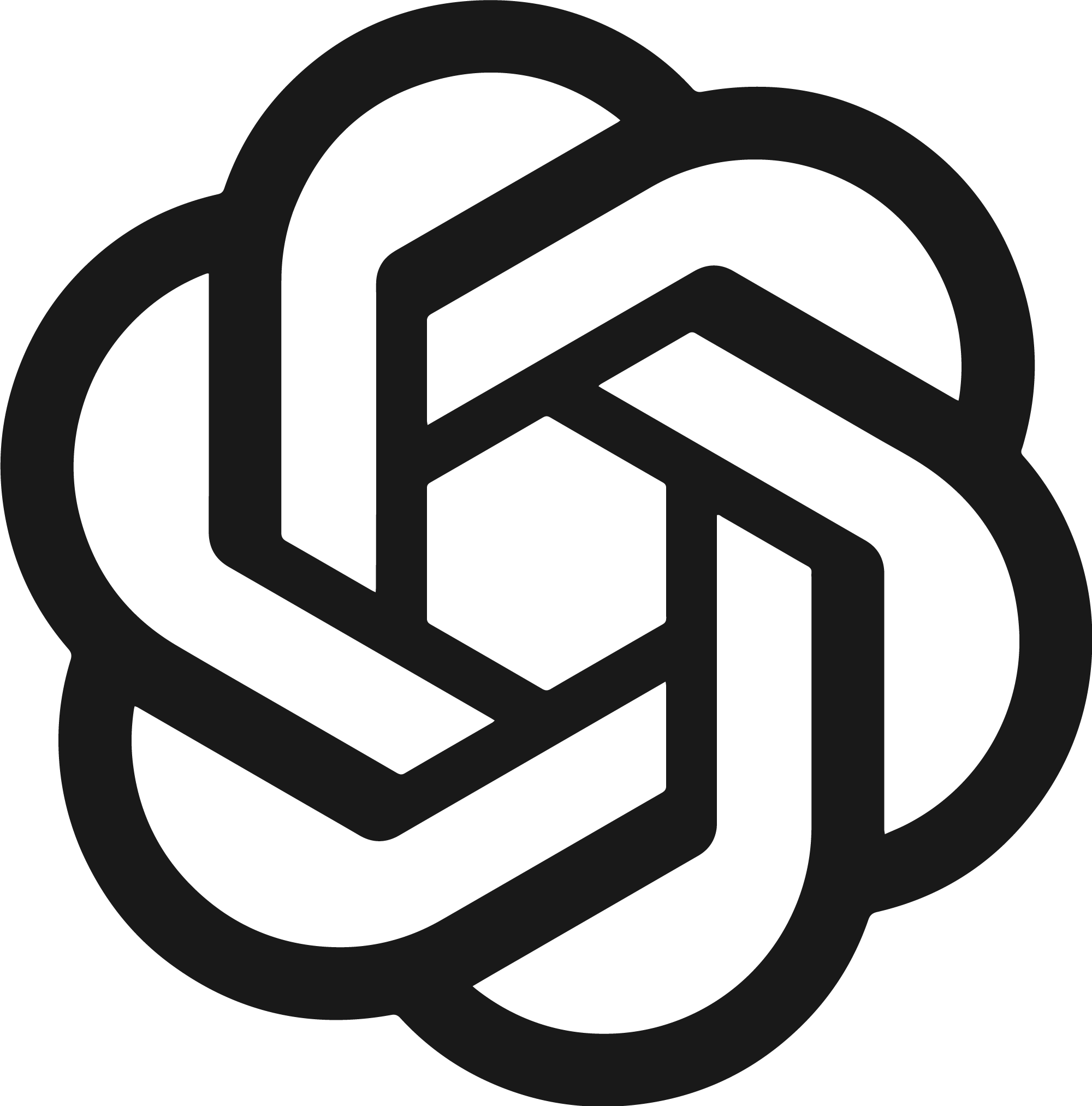}}

\usepackage{booktabs}
\usepackage{subcaption}
\usepackage{placeins} 

\begin{document}

\title[Benchmarking LLM-driven configuration repair]{Benchmarking LLM-Driven Network Configuration Repair}

\author{Ioannis Protogeros, Rufat Asadli, Benjamin Hoffman, Laurent Vanbever}
\affiliation{%
  \institution{ETH Zürich}
  \city{Zürich}
  \country{Switzerland}
}
\email{{iprotogeros, rasadli, bhoffman, lvanbever}@ethz.ch}

\renewcommand{\shortauthors}{Protogeros et al.}
\begin{abstract}
There is a rapidly growing interest in using Large Language Models (LLMs) to automate complex network operations, but their reliable adoption requires rigorous assessment of their effectiveness and safety. 
Existing benchmarks do not address whether LLMs can successfully resolve errors in large-scale, interdependent network configurations without introducing new disruptions. Developing such a benchmark is challenging: scenarios must be diverse and increasingly complex, yet their evaluation must be straightforward and meaningful.

In this paper, we present \textsc{Cornetto}, the first benchmark to evaluate LLM-driven network configuration repair \textit{functionally} and \textit{at scale}. \textsc{Cornetto} features a generation pipeline that synthesizes representative and plausible misconfiguration scenarios, coupled with an evaluation framework that uses formal verification to assess functional correctness of proposed fixes against ground-truth specifications.

Using this pipeline, we synthesize a dataset of 231 problems for fixing configurations across varying network topologies (20--754 nodes) and diverse protocols. We evaluate 9 state-of-the-art LLMs and find that while they show promise, they often introduce regressions and their performance degrades at scale. Our results indicate that reliable LLM-powered network automation requires integrating LLMs into iterative workflows guided by formal verification. 

%We open-source \textsc{Cornetto} at [anonymous link]

\end{abstract}

\maketitle

% include chapters
\section{Introduction}

\newcommand{\system}[0]{\textsc{Cornetto}\xspace}

% Possible structure for the intro

% Network correctness is paramount, yet hard to achieve.
% Network verification and synthesis have seen a lot of progress and success in the last ten years. Yet, these tools still struggle when it comes to usability and coverage.
% Recently, we have seen a surge of interests in using LLMs rather than models to reason about (mis)-configurations.
% LLMs tremendous potential, but there are also error- and hallucination-prone.
% Research question: How good are state-of-the-art (and future) LLMs at detecting and fixing misconfigurations?
% Answering this question requires a good benchmarks that are: ...
% A few benchmarks exist but they do not answer this question. Explain why
% We propose Cornetto, the first comprehensive benchmark for evaluating LLMs on misconfigurations...
% Key challenges of designing a good benchmarks: X, Y, Z.
% We address each challenge. For X, we do ... For Y, we do ..., ...
% We fully implemented Cornetto and use it to assess the perfomance of XXX LLMs of various sizes.
% Our finding indicate that: LLMs ...
% We release Cornetto open source
% To sum up, our contributions are:
% -
% -
% -
% Limitations: Cornetto asseses the performance of LLMs, it cannot XXX. This is still useful though as ... idea: The strategy space is huge so we don't evaluate everything in this paper. Still, Cornetto _can_ evaluate everything and can provide insights in yadda yadda and also be extended with the data generation pipeline (also potentially creating more difficult problems when needed?) 

Network correctness, while paramount, remains extremely difficult to achieve and maintain. While network verification and synthesis~\cite{batfish,minesweeper,veriflow,nsdi2018netcomplete,propane} have made significant strides towards eliminating human-induced misconfigurations, they are not a silver bullet. In particular, their adoption is hindered by limited protocol coverage and inaccuracies in modelling complex network behaviour~\cite{metha,modelfree-krentsel}.

%Given high-level objectives, operators must craft low-level configurations for large networks that run complex distributed protocols~\cite{propane}. This complexity leads to frequent misconfigurations, many of which can have disastrous consequences~\cite{meta_outage, minn_911, rogers_outage}.

More recently, there has been a surge of interest in leveraging Large Language Models (LLMs) as a more flexible approach to automating network operations. Hyperscalers have already begun deploying LLM-based frameworks that assist with such tasks, including ByteDance's NetAssistant~\cite{NetAssistant}, Alibaba's BiAn~\cite{BiAn}, and Meta's Confucius~\cite{wang2025confucius}. On the one hand, with their impressive capabilities across domains~\cite{gemini3report,gpt5systemcard}, LLMs appear promising for facilitating complex workflows that are bottlenecked by human reasoning. On the other hand, as probabilistic models, they remain prone to errors and hallucinations~\cite{stochastic-parrots,Ji_2023_hallucination}, precluding their adoption for managing critical infrastructure.

The tension between the tremendous potential for automating network operations and their associated risks makes principled evaluation indispensable. Doing so requires designing a benchmark that challenges LLM reasoning with complex and diverse configuration tasks, alongside with a proper methodology to assess their capabilities on said tasks. %but also provides meaningful evaluation based on achieved network correctness~\cite{continual-benchmarking}. 

Constructing such a diverse benchmark with plausible, well-posed problems is nontrivial. Unlike domains such as mathematics or software engineering, where benchmarks can leverage vast publicly available datasets, network configurations are proprietary and sensitive. Consequently, we must synthesize the dataset. Crucially, this synthesis cannot be random; to ensure relevance, test cases must reflect the complexities that operators face in production networks, including large-scale configurations with feature and protocol dependencies. Lastly, the evaluation must verify the functional correctness of solutions, \textit{scalably} and \textit{automatically}.

%It requires identifying and controlling the key dimensions that affect misconfiguration complexity to stress-test the models. At the same time, test cases must reflect real-world incidents, but relevant data (including network configurations) is proprietary and must therefore be synthesized. Lastly, the evaluation must verify the functional correctness of solutions, \textit{scalably} and \textit{automatically}.

\fakepar{The gap in LLM network configuration benchmarking} LLMs' progress outpaces our abilities to evaluate them effectively 
in fields that require deep domain expertise and complex reasoning. 
Despite growing interest in using LLMs for network operations, we still lack benchmarks to evaluate their performance and explore strategies to improve them.  

While previous works such as \textit{NetConfEval}~\cite{netconfeval} and \textit{Net\-LLMBench}~\cite{netllmbench} have established baselines for evaluating LLMs on network configuration tasks, they are severely constrained in scale and complexity relative to the capabilities of current models. Additionally, their proposed evaluation methods rely on proxy metrics (e.g., textual similarity or ping-test validation) that do not guarantee the functional correctness of a configuration. A more recent benchmark, NIKA~\cite{nika}, evaluates the diagnostic capabilities of LLM Agents in dynamic, emulated network environments. Yet, it does not support the evaluation of proposed fixes for network faults. 

Consequently, evaluating the performance of LLMs in repairing realistic, large-scale network configurations \textit{correctly} and \textit{safely} remains underexplored.

%\textcolor{red}{A common design limitation across the configuration-related aforementioned benchmarks is that they base their evaluation on ad hoc, problem-specific unit tests that validate solutions. While effective, this limits the generalisation of the benchmarking framework to new problem instances and doesn't allow reasoning about global effects beyond the defined tests.}

%but they often focus on limited scenarios or lack comprehensive evaluation metrics. \textcolor{red}{Expand on relevant benchmarks (NetConfEval   , NIKA, NetLLMBench,) and systems like NetAssistant, Confucius that are evaluated with ad-hoc, closed benchmarks}

\fakepar{\system: Correct \& safe configuration repair} To address this gap, we introduce \system, a benchmark that evaluates end-to-end configuration repair in representative, large-scale networks.\footnote{The idea behind the system was presented in a previously-accepted poster~\cite{protogeroscontinualbenchmarking}.} The task of resolving misconfigurations encapsulates critical challenges in network operations: understanding the interplay among interdependent features and protocols, and bridging the semantic gap between low-level configurations and high-level intent~\cite{propane}. To capture this complexity, \system employs a generation pipeline that produces syntactically valid configurations subject to logical and semantic constraints, thereby ensuring structural coherence and consistency. We use this pipeline to synthesize 231 challenging misconfiguration scenarios spanning diverse protocols and scales, where the intended network state is unambiguously defined. 

To evaluate correctness and safety, \system formally verifies the data plane of the reconfigured network against ground-truth specifications. This ensures that success depends on \textit{functional correctness}, requiring restoration of intended behaviour. The introduction of new bugs is penalised based on the extent of disruption to previously functional behaviour. Crucially, we also evaluate the diagnostic reasoning (localization and root-cause analysis) that led to the fix.

\fakepar{Key findings} Our evaluation of 9 LLMs reveals that while current models can diagnose and fix faults (restoring up to 60\% of network state on average), they cannot reliably act as monolithic solvers---the best-performing model successfully resolved only 25\% of scenarios. We find that performance degrades in large-scale settings with noisy data, and models often propose partial or unsafe solutions. These insights indicate that reliable automation requires integrating LLMs into systems that filter noisy data, preserve necessary context, and iteratively verify the safety of repairs before application. 

%We evaluate 9 state-of-the-art LLMs of various sizes, including flagship reasoning models like \textsc{GPT-5.2, Gemini 3 Pro} and \textsc{Claude 4.5 Opus}. Our findings show that while models show promise \textemdash~with the best-performing model restoring on average 60\% of the intended network state \textemdash~they struggle to fully resolve misconfigurations and often introduce new bugs. Even the top models successfully fix all faults in only 25\% of scenarios, while they induce from 3.6\% up to 14.1\% regressions on average. 

%\textcolor{red}{Some references to the ML Benchmarking book would be nice here}

\fakepar{Key contributions} Our summarized contributions are:
\begin{itemize}
\item We formulate the problem of automated configuration repair, enabling the quantification of functional disruption caused by misconfigurations and the assessment of fix correctness and safety.
\item We develop a scenario generation pipeline that synthesizes logically valid network configurations for any topology and systematically injects faults across diverse protocols.
\item Using this pipeline, we curate a dataset of 231 misconfiguration scenarios of varying scale and complexity by optimizing fault diversity within a minimal number of scenarios.
\item We design a verification-based evaluation pipeline that utilizes data-plane analysis to automatically infer ground-truth specifications and assess the functional correctness and safety of fixes.
\item We evaluate 9 state-of-the-art LLMs on \system and comprehensively analyze their performance, examining the effects of scenario complexity on reconfiguration correctness and safety.
\end{itemize}

\fakepar{Outlook} We will open source \system to the community as an extensible and modular framework.~\footnote{available at \url{https://github.com/nsg-ethz/cornetto}} \system enables the generation of more challenging tasks, and its architecture supports the evaluation of any LLM-based system, including advanced scaffolds such as Retrieval-Augmented Generation (RAG) systems and agentic setups. By providing a standardized testbed, \system contributes to the continual evaluation of LLMs on configuration repair.

\section{Overview}

\system is a benchmark for assessing LLMs' capabilities in automated network configuration repair. A \system \textbf{scenario} simulates an end-to-end troubleshooting task: Given a misconfigured network state (including topology and configuration files) and a set of high-level intents (e.g., Reachability between A and B), the model must localize and diagnose the misconfiguration, and propose a correct reconfiguration that restores the network's intended function.

\begin{figure*}[htbp]
    \centering
    \includegraphics[width=0.99\linewidth]{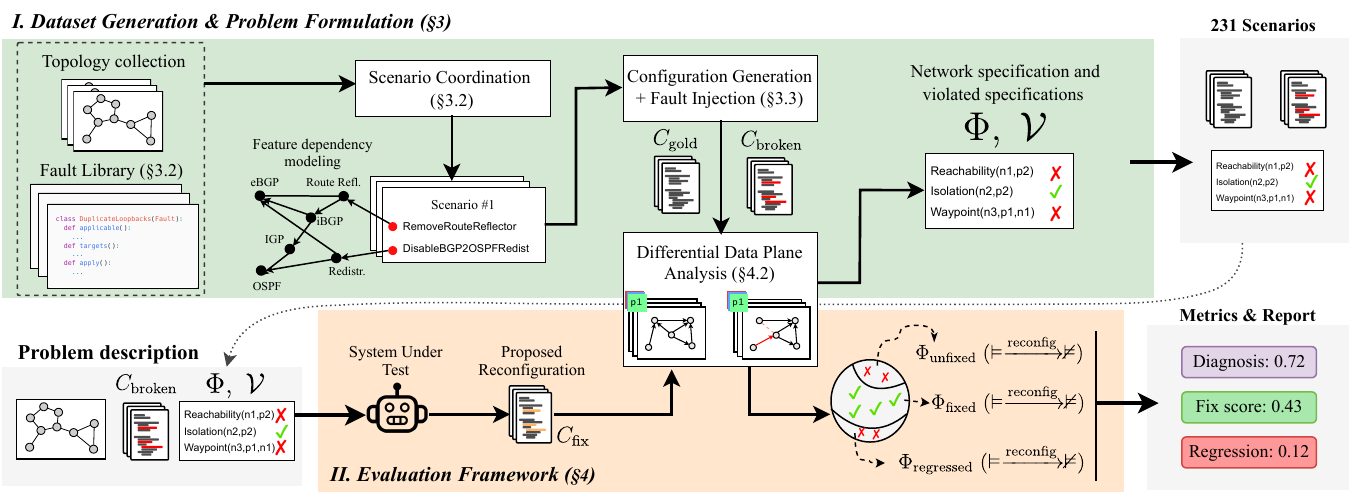}
    \caption{\system architecture. (I) The Dataset Generation pipeline coordinates the scenarios to ensure a \textit{diverse} and \textit{complex} test suite, generates \textit{sensible} configurations and misconfigurations, and provides a \textit{standardized} problem definition. (II) The Evaluation Framework enables automated and \textit{meaningful} evaluation of the created scenarios by validating the reconfigured network's behaviour against ground-truth specifications. }
    \label{fig:cornetto}
\end{figure*}

%\textcolor{red}{what is the system? what constitutes a scenario? how do we evaluate the different parts of troubleshooting: locate->diagnose->fix-refer to cisco troubleshooting or Hamadanian et al.? (key insight)}

\system comprises two pipelines. First, the \textbf{\textit{Dataset Generation}} focuses on creating realistic, diverse, and complex misconfiguration scenarios. Second, the \textbf{\textit{Evaluation Framework}} provides an automated system that enables meaningful evaluation of proposed solutions against intended network behaviour.

This section outlines \system design goals, the key insights required to address emerging challenges, and the system components that implement these solutions.

\subsection{Design Goals} 
%We establish three core design goals to ensure that \system provides value to the research community.
%\smallskip

%\noindent \textbf{\textit{I. Dataset Generation: Representative scenarios}}

\fakepar{Diversity and Complexity} To gain meaningful insights into the performance of LLMs, 
the benchmark must cover a wide variety of scenarios. 
This includes growing topology scales, 
features across diverse protocols, and complex fault scenarios
that reflect real-world issues. At the same time, the benchmark should remain small enough to keep its execution feasible without incurring high costs.

Additionally, the complexity of these scenarios should be sufficient to challenge models' reasoning capabilities and ensure resilience against benchmark saturation~\cite{hardt2025emerging}. A key challenge lies in identifying the dimensions that affect a scenario's complexity.

\fakepar{Sensible problems} Scenarios must be sensible so that a model's performance reliably proxies its real-world applicability. However, actual network-wide configurations and historical bug data are proprietary and hence unavailable. Therefore, the system must generate scenarios synthetically. This includes base configurations that are syntactically and semantically valid, maintain internal coherence, and respect dependencies between protocols. 
%\bigskip

\fakepar{Well-posed problems}  Despite the scenarios' complexity, the evaluation of their proposed solutions 
must be straightforward and provide concrete, interpretable metrics for success, efficiency and safety. This requires the concrete formulation of a misconfiguration scenario, including the definition and quantification of success metrics. 

%\noindent \textbf{\textit{II. Evaluation framework: Automatic and rigorous assesment}}

\fakepar{Meaningful evaluation}
%It is important to reason about whether a proposed fix actually resolves the misconfiguration
%without introducing new issues and specifying the process overhead. 
Evaluation should take into account the emergent network behaviour, 
rather than relying solely on the comparison of the configuration files. This 
requires analyzing the network's data plane behaviour post-fix, and comparing it
against a specification that captures the intended behaviour. Furthermore, all the different facets of troubleshooting should be assessed, including localization and diagnosis abilities.

%\begin{itemize}
%\item Meaningful and difficult yet easily verifiable problems.
%\item A possible approach: Use Batfish to calculate Specifications that change with introduction of each fault, and create \textit{pass} $\\rightarrow$ \textit{fail} ``unit tests". Those will comprise the \textit{observed symptoms} that will be part of the problem description we feed into the LLM. %reachability/waypointing, isolation, ordered preference, optional ECMP.
%\item Main indicator of success will be \% of unit tests (specifications) solved.
%\item We also measure \textbf{side-effects} as a safety indicator: change in next-hops, reachability/path length, table size, and POSSIBLY raw FIBs/RIBs. +Batfish's Differential Reachability
%\end{itemize}

\subsection{Key Insights}

%A main challenge lies in the tension between the open-ended nature of the problem
%space and the need for well-posed problem stipulations with verifiable solutions 
%and concrete yet interpretable metrics. We centre our design around the following insights:

\fakepar{\textit{Efficiently representing a massive task space} \mdseries \textit{(Enabling diversity)}} The space of all possible network configurations and misconfigurations is prohibitively large. To create a representative benchmark that can be run on a minimal compute budget, we identify the key dimensions that affect scenario difficulty (e.g., topology size, fault types) and employ sampling strategies and combinatorial testing techniques~\cite{kuhn2004-interaction-testing} to efficiently cover them. This enables \system to test LLMs against a diverse suite of scenarios without requiring thousands of redundant test cases.

\fakepar{\textit{Producing sensible configurations with grammar-based generation and semantic constraints} \mdseries \textit{(Providing sensible problems)}} For a benchmark to effectively evaluate LLM performance on resolving misconfigurations, it must contain \textit{sensible} configurations that can exist in reality. This means that configurations must be syntactically and semantically valid and realize a plausible intent. To achieve that, we first define a high-level logical network plan that specifies its functionality. Then, we synthesize configs using a grammar-based approach and enable features iteratively and contextually. This approach ensures that semantic constraints (e.g., that a prefix list is defined before it is referenced) and feature dependencies are respected. 

\fakepar{\textit{Formulating concrete problems through differential data plane analysis} \mdseries \textit{(Providing well-posed problems)}} We cannot proxy the effect of a misconfiguration by a textual diff; a perturbation of just a few Lines of Code (LoC) may cause significant disruptions in the network, while a larger change can be functionally benign. To rigorously define the problem, \system employs differential data-plane analysis. We compare the forwarding behaviour of the faulty network against the ``golden" reference state using Batfish~\cite{batfish}. The resulting set of behavioural differences (e.g., ``A cannot reach B") yields a concrete, symptom-based problem description.

%We perform data-plane analysis using the Batfish verifier~\cite{batfish} to extract the network's forwarding behaviour and compare it with the original, "golden" state, thereby quantifying the magnitude of the disruption. 

\fakepar{\textit{Evaluating functional correctness and reasoning} \mdseries \textit{(Enabling meaningful evaluation)}} To automate evaluation, we treat the network's functional requirements as a suite of ``unit tests". 
We mine the specific data-plane properties (Reachability, Isolation, Waypointing) satisfied by the golden network~\cite{birkner2020config2spec} to create a ground-truth specification. A correct solution will produce a reconfigured network that restores the previously violated predicates (i.e., it is \textbf{efficacious}), without violating any previously satisfied predicates (i.e., it is \textbf{safe}). Crucially, we also evaluate the intermediate steps of a model's reasoning (localization, root-cause diagnosis) to gain a holistic view into its troubleshooting effectiveness.

\subsection{\system}

\system comprises two primary pipelines, as shown in Fig.~\ref{fig:cornetto}. The \textbf{Dataset Generation (\S 3)} pipeline accepts a topology collection and a fault library to generate a minimal yet diverse suite of misconfiguration scenarios, along with their problem formulations. With the generated scenarios, the \textbf{Evaluation Framework (\S 4)} assesses repair capabilities by verifying configurations against a ground-truth specification.

%\fakepar{System inputs} The input to the dataset generation pipeline is 1) the set of topologies that will be used to generate the configurations (in our case, we use real-world topologies from the Internet Topology Zoo) and 2) A pre-defined library of faults, where each fault is a function that slightly perturbs a configuration in a way that breaks or alters the functionality of a protocol in the network. 

\fakepar{Scenario coordination and creation (\S\ref{subsec:space-representation})} To drive fault generation, we implement a diverse \textit{fault library}, where each fault is a function that slightly perturbs a configuration to break or alter the functionality of a protocol. The \textit{scenario coordinator} utilizes this library and the topology collection to orchestrate the creation of diverse scenarios, ensuring representation across all topology scales and fault combinations. For each scenario, the system generates a valid configuration that enables the specific protocols and features targeted by the fault. This process yields two network states per scenario: the \textit{Golden} (healthy) state and the \textit{Broken} (faulty) states.

\fakepar{Data plane analysis and problem formulation (\S\ref{subsec:scenario-generation})} The system invokes \emph{Batfish} to simulate the forwarding behaviour for both the Golden and Broken states. It then distills the behaviour of the Golden network into a set of invariants, or \textit{predicates}, which constitute the ground-truth specification. By verifying the Broken network state against this specification, the system identifies which specifications are violated. These violations are the ``symptoms" that indicate the problem in network behaviour.

\fakepar{Benchmark testbed (\S\ref{sec:eval-framework})} The testbed manages the interaction with the LLM-based system under test. The benchmarked system receives a description of the network problem created by the \textbf{\textit{Dataset Generation}} pipeline. This description contains the network topology, the faulty configurations, and the violated specifications (symptoms). \system tasks the model under test with a standard troubleshooting workflow: \textit{localizing} the fault, \textit{diagnosing} the root cause, and proposing a \textit{reconfiguration} to restore intended behaviour. 

The evaluation pipeline parses the proposed solution, simulates the new data plane, and compares the network state against the desired specification set. This process yields metrics for reconfiguration Efficacy (did the proposed reconfiguration fix the violations?) and Safety (did it introduce any new violations?), along with performance metrics for diagnosis and localization.

\fakepar{Output and results} For each test case, \system generates a structured report primarily containing:

\begin{itemize}
    \item Fix Rate (Efficacy): The proportion of initially violated specifications that are successfully restored by the configuration
    \item Regression Rate (Safety): Quantifies unintended side effects by measuring \textit{new} violations introduced by the reconfiguration.
\end{itemize}
Additionally, the testbed evaluates diagnostic quality using both objective metrics (precision/recall on faulty devices for localization) and an LLM-as-a-Judge~\cite{zheng2023judgingllmasajudgemtbenchchatbot} approach to assess provided textual diagnoses against ground-truth misconfigurations.

\section{Dataset Generation Pipeline}

\newcommand{\config}[1]{C_\text{#1}\xspace}
\newcommand{\router}[0]{\texttt{r}\xspace}
\newcommand{\prefix}[0]{\texttt{p}\xspace}

The test suite of our benchmark must cover a wide variety of high-quality scenarios of varying complexity. Yet, it should include a minimal number of test cases so the research community can test their methods without incurring prohibitive LLM inference costs. In this section, we define the network configuration troubleshooting task space and show how to represent it effectively, constructing a benchmark that meets our design goals.

\subsection{Task Definition}

Formally, we define a \system benchmark scenario as a tuple $(T, \config{gold}, \config{broken}, \Phi)$, where:
\begin{itemize}
    \item $T$ represents the network topology (undirected graph of devices and links between interfaces).
    \item $\config{gold}$ is the configuration at its ``Golden" state.
    \item $\config{broken}$ is the faulty configuration that is derived from $\config{gold}$ by applying a fault function $f$, so $\config{broken} = f(\config{gold})$.
    \item $\Phi$ is the set of data plane specifications (or intent) that the Golden Configuration satisfies. We denote satisfaction as $\config{gold} \models \Phi$. Since the forwarding plane of $\config{broken}$ deviates from intended behaviour, it holds that $\config{broken} \not\models \Phi$.
\end{itemize}

%\textcolor{red}{Maybe also formally define specification sets and specifications here?}

\subsubsection*{Specifications} We define a specification $\phi \in \Phi$ as a boolean predicate that describes a property in the network's forwarding behaviour. We consider four types of predicates that encompass the most common requirements about the network's function~\cite{birkner2020config2spec}:

\begin{itemize}
    \item \texttt{Reachability(r,p)}: Traffic from router \router can reach prefix \prefix.
    \item \texttt{Isolation(r,p)}: Traffic from \router cannot reach \prefix.
    \item \texttt{Waypointing(r,p,w)}: Traffic from \router destined to \prefix always passes through router \texttt{w}.
    \item \texttt{LoadBalancing(r,p,n)}: Traffic from \router destined to \prefix is load-balanced across \texttt{n} paths.
\end{itemize}

\subsubsection*{Specification violations} The fault $f$ introduces a disruption in the data plane. We define the set $\mathcal{V}$ of violated specifications as the subset of specifications that are satisfied by $\config{gold}$ \text{but violated by} $\config{broken}$: $$\mathcal{V} = \{\phi \in \Phi \mid \config{broken} \not\models \phi\}$$ 

\subsubsection*{The objective}

 The system under test's task is twofold. Given the input tuple $\mathcal{I}=(T,\config{broken},\mathcal{V})$, it must:
 
 \begin{itemize}
     \item \textbf{Localize and Diagnose:} Provide (i) a list of the misconfigured routers and (ii) a textual description of the faults in the network that correspond to the misconfiguration $\Delta(\config{gold}, \allowbreak\config{broken})$.
     \item \textbf{Repair:} Act as a repair function $\mathcal{R}$ that produces a reconfiguration $\config{fix} = \mathcal{R}(\mathcal{I})$ such that $\config{fix} \models \Phi$.
 \end{itemize}

%\subsubsection*{Repair performance metrics} Since a proposed solution can be partially correct, we score the reconfiguration $\config{fix}$ based on the restoration of intended behaviour \textcolor{red}{(might need to put this in the next chapter instead)}:

\subsection{Effective Task Space Representation}
\label{subsec:space-representation}

For a given topology set $\mathcal{T}$, the theoretical space of benchmark scenarios is defined by all the configurations that could take the place of $\config{gold}$ and $\config{broken}$. If we consider both configs to be a part of a configuration space $\mathcal{C}$ that fits each topology, then the scenario space would be contained in $\mathcal{T} \times \mathcal{C} \times \mathcal{C}$. This space is prohibitively large and dominated by unreasonable elements, i.e., random configuration pairs that cannot represent either operational networks or realistic misconfiguration scenarios. To construct a useful benchmark, we must restrict the space to a meaningful subset of scenarios and strategically represent it with minimal samples.
% This space, in addition to being prohibitively large, is also largely irrelevant because it includes many scenarios that would not make sense for the benchmark. 

\fakepar{Benchmark diversity} Covering the entire task space is neither possible nor relevant to our goals. However, to ensure robust evaluation, the benchmark should challenge the tested systems across different complexity dimensions.  

We identify the following controllable characteristics that are expected to critically affect the difficulty and nature of a misconfiguration scenario:

\begin{itemize}
    \item \textit{Topology scale}: Varying the size of the network up to hundreds of nodes will
    stress-test the models' ability to handle large inputs and identify the information that points to the issue.
    \item \textit{Number of applied faults}: Applying multiple simultaneous faults will challenge models with having to detect multiple independent root causes, potentially causing masking or compounding symptoms.
    \item \textit{Fault types}: Different faults from the library $\mathcal{F}$ will cause different types of symptoms that will be more or less difficult to link to the root cause.
\end{itemize}
These three dimensions will be used to select the scenarios that will comprise the benchmark. 

\fakepar{The fault library} To ensure the benchmark contains troubleshooting tasks across a wide array of misconfigurations, we curated a collection $\mathcal{F}$ of 27 fault functions that target specific protocol functionalities. This fault library spans across the following dimensions that we expect to affect scenario complexity:

\begin{itemize}
    \item \textit{Protocols and features affected}: The library includes faults that target features of eBGP, iBGP (incl. route reflection), OSPF and IS-IS (single and multi area), redistribution, ACLs, route-maps, and static routes
    \item \textit{Configuration impact}: From perturbing a single parameter (e.g., the subnet mask of an interface) to performing ``organized" alterations like removing a route reflector functionality from a router.  
    \item \textit{Operational impact}: Faults that disrupt a protocol's functionality (e.g., mismatched \texttt{remote-as} numbers preventing BGP session establishment), and faults that only change the intent of the used feature (e.g., stripping an export policy) 
\end{itemize}
\noindent Appendix~\ref{appendix:fault-library} contains the comprehensive list of all faults.

%\textcolor{red}{describe the faults in the library, how they contribute towards diversity (various protocols, nature of faults, why was this chosen instead of Grammar based fuzzing? (targeted protocol disruptions, syntactic mistakes can be mostly too easy to detect, random fuzzing would generally not lead to interesting scenarios))}

\fakepar{Representative Sampling Strategy} To balance scenario diversity with a manageable test set size, we employ a sampling strategy centred on \textit{pairwise coverage}~\cite{kuhn2004-interaction-testing}. Our goal is to generate complex scenarios with up to $N=8$ simultaneous faults, where every possible pair of fault types appears together at least once. This ensures we test model performance on scenarios with varying disruptions and multi-root-cause failures.
%To balance our goals of creating as diverse scenarios as possible, making sure they can represent plausible scenarios, and having our test set stay within a sensible size, we sample the task space using a strategy centered around the following points:
% \begin{itemize}
%     \item We limit the maximum number of simultaneous faults to $N=8$, ensuring scenarios range from ``simpler" faults to more disruptive multi-root-cause failures.
%     \item We optimize the scenario set for \textit{pairwise coverage}~\cite{kuhn2004-interaction-testing}, ensuring that every possible pair of fault types appears together at least once. This maximizes the diversity of possible fault interactions while minimizing the total number of scenarios.
%     \item We represent all topological scales by applying each fault pattern to small-, medium-, and large-scale network topologies.
% \end{itemize}

\subsubsection*{1. Fault selection procedure} We construct a compact collection of fault sets $\mathcal{S}$ by iteratively sampling from the fault library $\mathcal{F}$ until 100\% pairwise coverage is achieved. For each generation step:

%Defining each benchmark scenario starts with selecting the set of faults to apply. To ensure diversity in incident complexity and the concurrent types of applied faults, while minimizing the number of scenarios, we employ the following procedure that maximizes \textit{pairwise coverage} across fault types. For each generation step:

\begin{enumerate}[leftmargin=*]
    \item We randomly select a number $k \in \{2,\dots,8\}$ of simultaneous faults to apply. This variation ensures the benchmark includes scenarios from few to many root causes.
    \item Select a subset $F_s \subset \mathcal{F}$ of $k$ faults $|F_s|=k$ that greedily maximizes the number of \textbf{newly covered fault pairs}
    \item Repeat this process until the set of applied scenarios covers 100\% of feasible fault pairs.
\end{enumerate}
This optimization yields a compact collection of $50$ distinct fault sets $\mathcal{S}=\{F_1,F_2,\dots F_{50}\}$ that describe which faults are to be applied in each scenario. To this collection, we add all monosets of single faults ($k=1$) to include each fault type individually.

\subsubsection*{2. Topology selection} For the topology collection $\mathcal{T}$, we use real-world topologies from the Topology Zoo~\cite{topologyzoo}. To ensure these fault patterns are tested across varying scales, we stratify our topology dataset $\mathcal{T}$ into three tiers: \textbf{Small} (<50 nodes), \textbf{Medium} (50--100 nodes), and \textbf{Large} (>100 nodes).

For each generated fault set $F_i \in \mathcal{S}$, we instantiate the benchmark scenario by applying the faults to three distinct topologies, one randomly sampled from each tier. This yields a final dataset of \textbf{231 scenarios} that cover diverse combinations of fault types across all scale tiers.

\subsection{Scenario Generation}
\label{subsec:scenario-generation}
The scenario selection procedure described before yields for each scenario a descriptor $(F,T)$ with (i) the set of faults $F \subset \mathcal{F}$ and (ii) the topology of the network $T \in \mathcal{T}$. We now describe the pipeline that synthesizes the configurations themselves and applies the fault functions to obtain the configurations $\config{gold}$ and $\config{broken}$.

To ensure the benchmark scenarios are plausible and useful, the generated configurations must satisfy two constraints:

\begin{itemize}
    \item The base configuration $\config{gold}$ must be syntactically and semantically valid. Crucially, the features that are enabled in the network must not exist vacuously (e.g. BGP processes without peers or unreferenced route-maps), but they must be structured to realize a functional intent within the network context.
    \item The ``broken" config $\config{broken}$ must be derived from $\config{gold}$ after a perturbation that is \emph{minimal}, so it can represent a plausible misconfiguration.
\end{itemize}

Existing synthesizers like NetComplete~\cite{nsdi2018netcomplete} or Propane~\cite{propane,propaneAT} are ill-suited for this task because they optimize for a fundamentally different objective: finding \emph{any} valid configuration that satisfies a specific high-level intent, often resulting in simple, uniform implementations. Our benchmark requires configurations that vary in protocols and \textit{specific, low-level} configuration features. High-level intent is insufficient for this purpose, as a single intent (e.g., reachability) can be satisfied by many combinations of features. Therefore, we adopt a grammar-based generation approach building on \textit{Metha}~\cite{metha}, allowing for the selection of features that the fault functions in $\mathcal{F}$ can affect.

\subsubsection*{Configuration feature selection} Each fault function acts upon specific configuration features. But for the fault to be applicable, the appropriate "attack surface" must exist in the first place. For instance, removing a route reflector requires that the AS is configured with route reflection clusters.

To ensure that generated configurations are operationally viable, we model the dependencies between network protocols. Network features are rarely independent; for example, testing a route-reflection fault requires the AS to support iBGP, which, in turn, relies on an underlying IGP (e.g., OSPF) for loopback reachability. We enforce these constraints during the generation phase: for each selected fault, the generator enables the target features and recursively satisfies all its prerequisite protocol dependencies.

\subsubsection*{Configuration generation} The syntactic and semantic validity of the ``Golden" state configurations is critical for the realism and utility of the benchmark. While syntactic validity ensures the configuration can be parsed, semantic validity ensures the configuration is logically consistent. We enforce two types of semantic constraints, as stipulated in \textit{Metha}:

\begin{itemize}
    \item \textbf{Intra-device constraints:} Dependencies within a single configuration file. For example, a BGP neighbour statement cannot apply a \texttt{route-map} that has not been defined, and an interface cannot be assigned to an OSPF area if the OSPF process is not active.
    \item \textbf{Inter-device constraints:} Dependencies across the network. For example, two routers connected via a link must have IP addresses in the same subnet, and eBGP peers must have matching \texttt{remote-as} declarations.
\end{itemize}

To satisfy these, we construct a high-level \textit{logical plan} of the network. This process is iterative and context-aware: First, we extend the physical topology with logical groupings to define the control plane hierarchy. We split the topology into ASes and assign OSPF/IS-IS areas to router interfaces according to the chosen IGP in each domain. We also define peering relationships, including iBGP full-meshes and route-reflection clusters. Then, we assign subnets to links and IP addresses to the interfaces on those links. Finally, additional resources are generated based on the selected features and their dependencies, as defined by the logical topology. For instance, BGP advertisements are generated strictly for subnets assigned to the router's local interfaces.

Once the logical plan is defined, we render it using a template that follows a Context-Free Grammar for vendor-specific configurations (e.g., Cisco IOS). This part ensures the syntactical correctness of the produced configurations, and is decoupled from the process of enforcing semantic constraints (which are not context-free).

\subsubsection*{Fault injection} We apply the fault functions to the produced logical plan, so when it passes through the renderer, it results in the broken configuration $\config{broken}$. After these faults, the configurations may either violate semantic constraints (e.g., mismatched \texttt{remote-as} parameters) or remain semantically valid, only deviating from intended behaviour (e.g., changing a route-map action from permit to deny). In both cases, the specifications violated after fault injection are verified through data-plane analysis, ensuring that all faults induce a tangible change in forwarding behaviour. 

\subsection{Dataset Statistics} The resulting \system dataset comprises 231 network misconfiguration scenarios across topologies with 20 to 754 nodes. Table~\ref{tab:main_dataset_summary} reports the key features of the scenarios. Notably, the network-wide configurations are large (>16K lines of code on average), yet contain only a few lines actually affected by the fault. Consequently, solving \system requires navigating massive, distributed configuration files to locate the few relevant lines that point to an issue.

\begin{figure}[ht]
    \centering
    \includegraphics[width=0.99\linewidth]{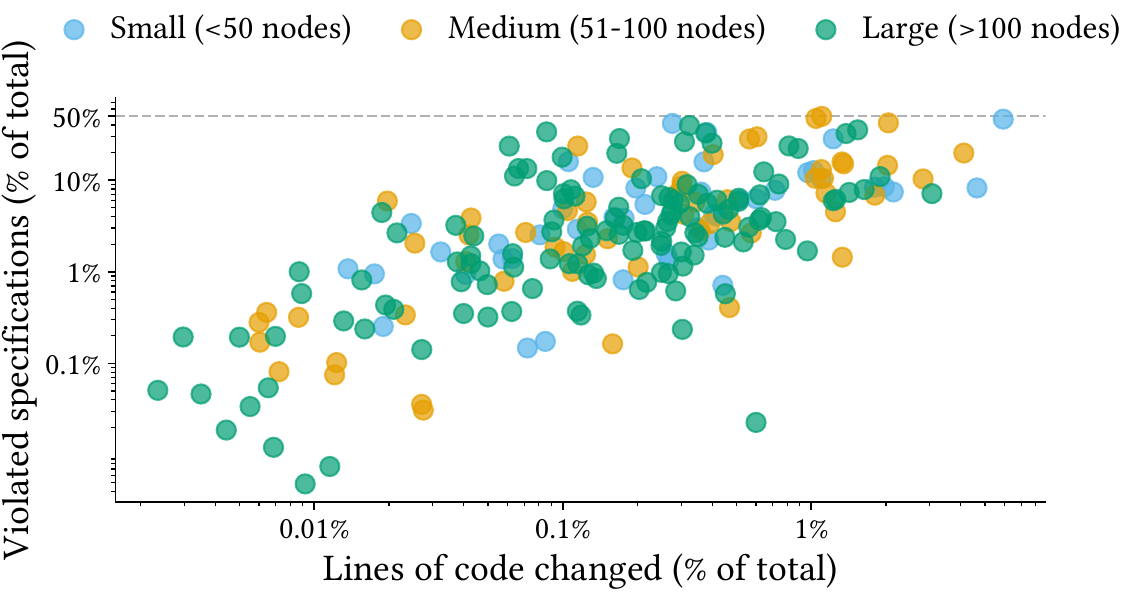}
    \caption{While configuration perturbations are minimal, disruption in network behaviour varies greatly across scenarios.}
    \label{fig:loc_vs_disruption}
\end{figure}

At the same time, the actual change in the configuration is small and buried under the volume of code. The effect of a misconfiguration is not related to its textual difference either; most of the faults affect <1\% of the total configuration lines, but may greatly affect the functionality of the network as shown in Fig.~\ref{fig:loc_vs_disruption}. We expect the varying levels of perturbations and symptoms to affect scenario difficulty.

\begin{table}[ht]
    \centering
    \caption{\system spans diverse scales and misconfiguration severities.}
    \label{tab:main_dataset_summary}
    \begin{tabular}{llrr}
    \toprule
    \textbf{} & \textbf{Metric} & \textbf{Mean} & \textbf{Max} \\
    \midrule
    Topology      & Nodes (\#)                    & 86.4   & 754  \\
                  & Configuration lines (LoC)                 & 16.1K  & 200.0K \\
                  & Routes               & 4.6K   & 130.4K \\
                  & Data plane predicates           & 12.7K  & 598.0K \\
    \addlinespace
    Fault impact & Lines edited                  & 50.5   & 345  \\
    &   Routers affected & 5.5 & 20 \\
                  & Routes changed                & 568.4  & 9.2K   \\
                  & Predicates changed            & 1.5K   & 38.4K  \\
%                  & Forwarding groups changed     & 82.1   & 651.0  \\
    \addlinespace
    Impact (\%)   & LoC\% changed                 & 0.44   & 5.93   \\
                  & Routes\% changed     & 6.34   & 57.3   \\
                  & Predicates\% changed  & 6.82   & 49.8   \\
    \bottomrule
    \end{tabular}
\end{table}

\section{Evaluation Framework}
\label{sec:eval-framework}

To derive meaningful insights into the diagnostic capabilities of LLMs, the evaluation system must distil concrete, interpretable metrics that quantify the functional success of a solution. In this section, we delineate the \system evaluation pipeline that enables automatic evaluation of proposed reconfigurations against the ground truth of the network's data-plane behaviour.

\subsection{LLM-Benchmark Interface}
\label{subsec:benchmark-interface}
To enable testing and comparing different models and systems around them, we build a standardized interface that decouples evaluation logic from the specific solver. Hence, \system can evaluate any proposed system that can generate a configuration $\config{fix}$. To carry out experiments across different models, we implement an evaluation framework that (i) builds a structured prompt containing the problem description to elicit a solution, and (ii) parses a model-generated patch to build the configuration $\config{fix}$. 

\subsubsection*{Context Construction} To provide the necessary info to diagnose and fix the problem in the configuration, the system constructs a prompt containing three main information sources:
\smallskip
\begin{itemize}
    \item The physical topology $T$
    \item The list of violated specifications $\mathcal{V}$
    \item The faulty configuration files $\config{broken}$
\end{itemize}

A key challenge for the benchmarked system is handling the volume and sparsity of the available raw data; among thousands of configuration lines, only a very small subset points to the root cause in the network. At the same time, it is possible that network-wide configurations, along with the specifications and topology data, cannot fit within the LLMs' context windows, and their performance is known to degrade well before that limit~\cite{shi2023largelanguagemodelseasily, liu2023lostmiddlelanguagemodels}.

Consequently, handling the context constitutes a core experimental dimension (\S\ref{sec:experimental-setup}). We define the system under test to include not only the generation model but also the \textit{context strategy}, that is used to derive, filter, or retrieve relevant information from the available raw data ($T$, $\config{broken}$, $\Phi$, $\mathcal{V}$).

For obtaining the solution, we prompt the models to output the following:
\smallskip
\begin{enumerate}
    \item A textual diagnosis, containing all the detected faults in the network configuration $\config{broken}$
    \item A list of all the routers that need to be reconfigured, and the needed reconfigurations to resolve the specification violations
\end{enumerate}

\subsubsection*{Reconfiguration Parser} The raw text output of the model needs to be processed in order to apply the fixes and obtain the proposed reconfiguration $\config{fix}$. We decided against requiring Unix diff patches~\cite{gnu_diffutils}, since they require precise line arithmetic, which language models famously struggle with~\cite{glukhov2025diffxyzbenchmarkevaluatingdiff, swebench}. Instead, following common practice in coding agents~\cite{aider} that perform edit operations, we expect the answer to include, for each reconfigured file:
\begin{itemize}
    \item A \textit{search block}, containing a snippet that is \textbf{uniquely present} in the configuration file
    \item A \textit{replace block}, containing the snippet that will replace the search block
\end{itemize}

Since unparseable output is still possible (wrong format, non-existent search block), we draw inspiration from the same practices and robustify the pipeline by \textit{(i)} allowing fuzzy matching of \textit{search blocks}, if there is a block that differs only in whitespaces or has a small enough \textit{Levenshtein distance}~\cite{levenshtein1966binary}, and \textit{(ii)} by providing feedback from the parser in case of invalid outputs.  

\subsection{Differential Data Plane Analysis}
\label{subsec:data-plane-analysis}
Evaluating the functional correctness of the proposed reconfiguration $\config{fix}$ requires extracting the network's emergent high-level behaviour from its low-level configuration files and comparing it against a ``ground truth" behaviour. 

\begin{figure}[t]
\centering
\resizebox{\columnwidth}{!}{%
\begin{tikzpicture}[
    % Increased spacing between the main stages to prevent overlap
    node distance=0.8cm and 1.0cm,
    % Styles
    router/.style={circle, draw=black, fill=gray!10, thick, minimum size=0.25cm, font=\tiny\bfseries, inner sep=1pt},
    box/.style={rectangle, draw=black, thick, align=center, fill=white, font=\scriptsize, inner sep=4pt},
    arrow/.style={->, >=Stealth, thick},
    badarrow/.style={-, >=Stealth, thick, red, dashed},
    flowarrow/.style={->, >=Stealth, ultra thick, gray!60}
]

% --- 1. LEFT: Rules Table (No Title) ---
\node[box] (table) {
    \begin{tabular}{l|l|l}
    \textbf{Node} & \textbf{Dest.} & \textbf{Action} \\ \hline
    R1 & 10.1/24 & Fwd R2 \\
    R1 & 10.1/24 & Fwd R3 \\
    R2 & 10.1/24 & Fwd R4 \\
    R3 & 10.1/24 & Drop \\
    R4 & 10.1/24 & Accept \\
    \end{tabular}
};

\coordinate[right=of table, xshift=0.93cm] (graph_center);

\node[router] (r1) at ([xshift=-0.6cm]graph_center) {R1};
\node[router] (r2) at ([yshift=0.6cm]graph_center) {R2};
\node[router] (r3) at ([yshift=-0.6cm]graph_center) {R3};
\node[router] (r4) at ([xshift=0.6cm]graph_center) {R4};

% Edges
\draw[arrow] (r1) -- (r2);
\draw[arrow] (r1) -- (r3);
\draw[arrow] (r2) -- (r4);
\draw[badarrow] (r3)--(r4);

% Graph Container 
\node[draw=black!50, dashed, fit=(r1) (r2) (r3) (r4), inner sep=2pt] (graphbox) {};

% --- 3. RIGHT: Predicates ---
\node[box, right=of graphbox, align=left] (predicates) {
    %\tiny
    Reachability(R1,10.1/24) \\
    %\tiny
    Reachability(R2,10.1/24) \\
    %\tiny
    Isolation(R3,10.1/24) \\
    %\tiny
    Reachability(R4,10.1/24) \\
    %\tiny 
    Waypoint(R1,10.1/24,R2)
};

% --- PIPELINE ARROWS ---
\draw[flowarrow] (table.east) -- (graphbox.west);
\draw[flowarrow] (graphbox.east) -- (predicates.west);

\end{tikzpicture}
}
\caption{\emph{For each} unique destination prefix, the pipeline uses the forwarding behaviour table (calculated by Batfish) to construct a forwarding graph, from which it derives the specifications of the network.}
\label{fig:pred_extraction}
\end{figure}
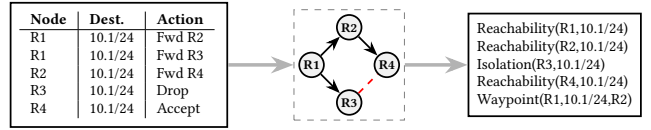

We standardize a network's high-level function with the following procedure: First, we use Batfish to simulate the data plane of each network, including forwarding decisions for each \textit{(src, dst prefix)} pair. 
Then, we construct a \textit{forwarding graph} for each prefix, and use the graph algorithms proposed in~\textit{Config2Spec} \cite{birkner2020config2spec} to extract the set of predicates that describe the specifications of the network. Fig.~\ref{fig:pred_extraction} shows an example of this flow. 

\fakepar{Specification-based reconfiguration evaluation} With the previous process we obtain from $\config{gold}$ the set of specifications $\Phi$ that the reconfigured network must satisfy. 
To evaluate a proposed reconfiguration, we also need to compare the behaviour between each network state.
We do this by extracting and comparing the high-level specifications across networks. To improve efficiency, we construct the forwarding graphs only for prefixes whose entries in the forwarding behaviour table differ from the golden state. We calculate the set of violated specifications $\mathcal{V}$, and after performing the predicate extraction pipeline for $\config{fix}$, we calculate the following sets on which we will base our scoring:

\begin{itemize}
    \item The set of successfully resolved violations 
    \[ \Phi_{\text{fixed}} = \{ \phi \in \mathcal{V} \mid \config{fix} \models \phi \} \]

    \item The set of regressions as originally healthy specifications that are violated by the fix \[ \Phi_{\text{regressed}} = \{ \phi \in (\Phi \setminus \mathcal{V}) \mid \config{fix} \not\models \phi \} \]

    \item The set of violations that remained unresolved: \[ \Phi_{\text{unfixed}} = \mathcal{V} \setminus \Phi_{\text{fixed}} \]

\end{itemize}

And we calculate the following scores that describe different performance aspects of the solutions:

\begin{itemize}

    \item \textbf{Safety (Regression Rate):} The proportion of specifications that were violated because of the proposed misconfiguration:
    \[ \text{Regression} = \frac{|\Phi_{\text{regressed}}|}{|\Phi_{\text{fixed}}|+|\Phi_{\text{unfixed}}| + |\Phi_{\text{regressed}}|} \]

    \item \textbf{Efficacy (Fix Score):} The proportion of resolved specifications relative to the total violations (initial and regressions):
    \[ \text{Fix Score} = \frac{|\Phi_{\text{fixed}}|}{|\Phi_{\text{fixed}}|+|\Phi_{\text{unfixed}}| + |\Phi_{\text{regressed}}|} \]
\end{itemize}

\subsection{Diagnosis Evaluator}
\label{subsec:judge-diagnosis}
While the previous procedure quantifies the restoration of the intended network behaviour, we are also interested in evaluating LLMs' ability to localize (at the router-level) and diagnose root causes. The following parts of a proposed solution are related to this:

\begin{itemize}
    \item The list of routers that are detected
    \item The textual description of the detected faults
\end{itemize}

We use these artifacts to gain insights into how \textbf{sound} (are \textit{only} real faults detected?) and how \textbf{complete} (are \emph{all} the faults detected) the proposed diagnoses are. For the task of localizing faulty routers, computing precision and recall is straightforward because they can be compared against the ground truth.

To evaluate diagnostic accuracy, a naïve approach would be to have the models classify each fault into a class from the fault library $\mathcal{F}$. However, this would require exposing the model to the list of potential fault types, thereby contaminating the reasoning process and compromising the generalizability of the open-ended diagnosis problem.

Instead, we use the LLM-as-a-Judge method~\cite{zheng2023judgingllmasajudgemtbenchchatbot}, as a viable and scalable alternative to expert human annotation, where we provide 3 different high-capability LLMs (\textsc{GPT-5.1, Claude 4.5 Opus, Gemini 2.5 Pro}) with (i) the proposed textual diagnosis of the benchmarked LLM, and (ii) the ground truth list of misconfigurations in the network; both the types of faults and the textual differences. We request two separate scores that quantify the completeness (i.e., the percentage of faults correctly identified) and the soundness (i.e., the percentage of faults hallucinated) of the diagnosis. To obtain the final scores, we aggregate scores across the judge models to further ensure robustness and mitigate potential biases.
\section{Experimental Setup}
\label{sec:experimental-setup}
%The proposed benchmark framework should effectively illustrate, through explainable metrics, how capable LLMs are to detect and resolve network misconfigurations. To this end, we extend the discussion to a detailed description of how the evaluation environment interacts with the selected LLMs. We also explain our efforts toward creating a unified evaluation algorithm, the results of which are compatible across different systems-under-test and can be easily interpreted with respect to the real-world network troubleshooting practices.
%\subsection{Selected models}

In this section, we review the models examined and describe how inputs are constructed to evaluate LLMs on \system.

\fakepar{Model selection} We evaluate 9 LLMs of varying sizes on \system: 8 state-of-the-art proprietary LLMs (including \textsc{GPT-5.2, Gemini 3.0}, and \textsc{Claude 4.5 Opus}) that consistently dominate benchmark leaderboards~\cite{swebench,matharena} and a single open-source model: \textsc{GPT-OSS-20B}. We include the latter to test the viability of smaller, lower-cost models, though we expect it to be outperformed by the larger proprietary ones.

\fakepar{Input and context construction} A model receives as input a topology $T$, a set of violated specifications $\mathcal{V}$, and a network-wide configuration $\config{broken}$, which is potentially very large. This data is noisy and voluminous, posing a distinct challenge for the LLM-based troubleshooting workflow~\cite{jiang2024caipdetectingroutermisconfigurations, holistic-incident-management}. For evaluating LLMs' robustness in handling this volume and noise, we test the following strategies for including configurations in the context:

\begin{itemize}
    \item \textbf{Full context}: The model receives the entire network-wide configuration $\config{broken}$. Because of the high context window limits of current models, the vast majority (98\%) of cases can fit entirely within the prompt. If a model cannot handle the entire input, configuration files are truncated.
    \item \textbf{Oracle context}: The model receives only the files affected by the misconfiguration. This is an idealized scenario for analysis purposes, since realistically, this information is not known a priori.
    \item \textbf{Retrieval mode}: For a subset of models, we evaluate a two-stage workflow where the LLM is first prompted to retrieve the necessary configuration files for diagnosis. This should yield a superset of the faulty configurations; therefore, we evaluate the success of this step using the \textit{recall} metric.
\end{itemize}

\fakepar{Prompting and parsing} Following standard practices in tackling holistic and multi-stage reasoning tasks~\cite{wei2023chainofthoughtpromptingelicitsreasoning, zhou2023leasttomostpromptingenablescomplex}, we design a prompt that decomposes the reconfiguration repair problem into a structured workflow. We instruct the model to use a \textit{Chain-of-Thought} (CoT) process that mirrors the formal fault management lifecycle standard~\cite{iso7498}.

 We implement a parser that expects three distinct solution parts that map to each goal of the process:
 \begin{itemize}
     \item \textbf{Localization}: The list of detected faulty routers.
     \item \textbf{Diagnosis}: A textual diagnosis of the root causes in the configuration. 
     \item \textbf{Reconfiguration}: A list of configuration changes for the detected faulty routers, using the mandated search-and-replace format (as described in~\S\ref{subsec:benchmark-interface}).
 \end{itemize}

This structured elicitation allows \system to evaluate the accuracy of the intermediate reasoning steps (localization and diagnosis) independently of the final repair quality. 
Finally, to ensure that format following is not a confounding factor in our evaluation, we configure the parser to allow one retry attempt per scenario if the model produces a solution that cannot be parsed.

\fakepar{Metrics and reporting} We evaluate performance for each stage leading up to the configuration repair.

\subsubsection*{1. Diagnostic Reasoning} 
To assess the models' ability to isolate faults before fixing them, we report two key metrics:
\begin{itemize}
    \item \textbf{Localization F1:} The harmonic mean of precision and recall for the set of faulty routers identified by the model against the ground truth.
    \item \textbf{Diagnosis Quality:} Using the LLM-as-a-Judge method (\S\ref{subsec:judge-diagnosis}), we quantify the \textit{soundness} and \textit{completeness} of the model's natural language explanation against the ground truth misconfigurations provided to the LLM-Judges. %\textcolor{red}{sound.+complt. and the overall subjective judge score no? A: judge score is an average of the two}
\end{itemize}

\subsubsection*{2. Functional Correctness} 
We use differential data plane analysis and report the \textit{fix rate} and \textit{regression rate} metrics for each scenario, as defined in \S\ref{subsec:data-plane-analysis}. We also report the percentage of cases correctly resolved, i.e., fixes that satisfy the intended specification without introducing new regressions.

\begin{table*}[!ht]
\centering
\begin{tabular}{lcccccc}
\toprule
\textbf{Model} & \textbf{Fix Score} $\uparrow$ & \textbf{Localization} $\uparrow$ & \textbf{Diagnosis} $\uparrow$ & \textbf{Regression} $\downarrow$ & \textbf{Success Rate} $\uparrow$ & \textbf{Cost (\$/task)} $\downarrow$ \\
\midrule
\textsc{GPT-5.2 (High)} & 57.8 & 76.5 & 76.8 & 8.6 & 25.5 & 0.16 \\
\textsc{Gemini 3 Flash} & 55.4 & 73.7 & 70.2 & 11.3 & 24.2 & 0.04 \\
\textsc{Gemini 3 Pro}   & 47.2 & 70.9 & 66.7 & 13.9 & 18.6 &  0.18 \\
\textsc{GPT-5.1 (High)} & 45.4 & 60.7 & 65.6 & 5.3 & 22.9 &  0.11 \\
\textsc{Claude 4.5 Opus} & 44.2 & 59.8 & 64.1 & 3.5 & 24.3 &  0.42 \\
\textsc{Claude 4.5 Sonnet} & 37.2 & 58.1 & 60.6 & 5.4 & 17.3 &  0.28 \\
\textsc{GPT-5 mini (High)} & 33.9 & 57.5 & 58.3 & 12.8 & 16.9 &  0.02 \\
\textsc{Grok 4.1 Fast (R.)} & 4.5 & 17.1 & 45.2 & 4.9 & 0.03 &  0.01 \\
\textsc{GPT-OSS-20B}        & 1.7 & 15.2 & 12.5 & 2.0 & 0.01 & - \\
\bottomrule
\end{tabular}
\caption{While models show promise in restoring network state, they rarely achieve complete resolution and frequently introduce regressions.}
\label{tab:main_leaderboard_table}
\end{table*}

\section{Evaluation}
\label{sec:evaluation}

We use \system to evaluate 9 state-of-the-art LLMs across 231 diverse troubleshooting scenarios. Through our analysis, we aim to answer two primary research questions:

\begin{itemize}[leftmargin=*]
    \item \textbf{RQ1:} \textit{To what extent can current LLMs autonomously localize, diagnose, and repair network misconfigurations without introducing regressions?} We present the performance of all LLMs across our core metrics for diagnostic accuracy and repair functional correctness, summarizing key insights into their capabilities.
    \item \textbf{RQ2:} \textit{How do task factors such as topological and configuration scale, fault multiplicity, and extent of disruption impact model performance}? We quantify the degradation of model reliability across increasing difficulty gradients.
\end{itemize}

%We provide all experimental results in this section as we evaluate LLMs across various settings. We first evaluate all 9 models on 231 tasks and present their performance across core metrics (\S\ref{subsec:main_results}). Later, we analyze in detail how LLMs' performance shifts with increasing task difficulty gradients (e.g., topology/input prompt size, number of faults injected/routers affected) (\S\ref{subsec:task_complexity}). We further extend this analysis to discuss the potential effects of different context window filling strategies (\S\ref{subsec:context_handling}).

% ------------ LEADERBOARD MAIN SCORE PLOT ------------
\begin{figure}[h]
    \centering
    \includegraphics[width=1.01\linewidth]{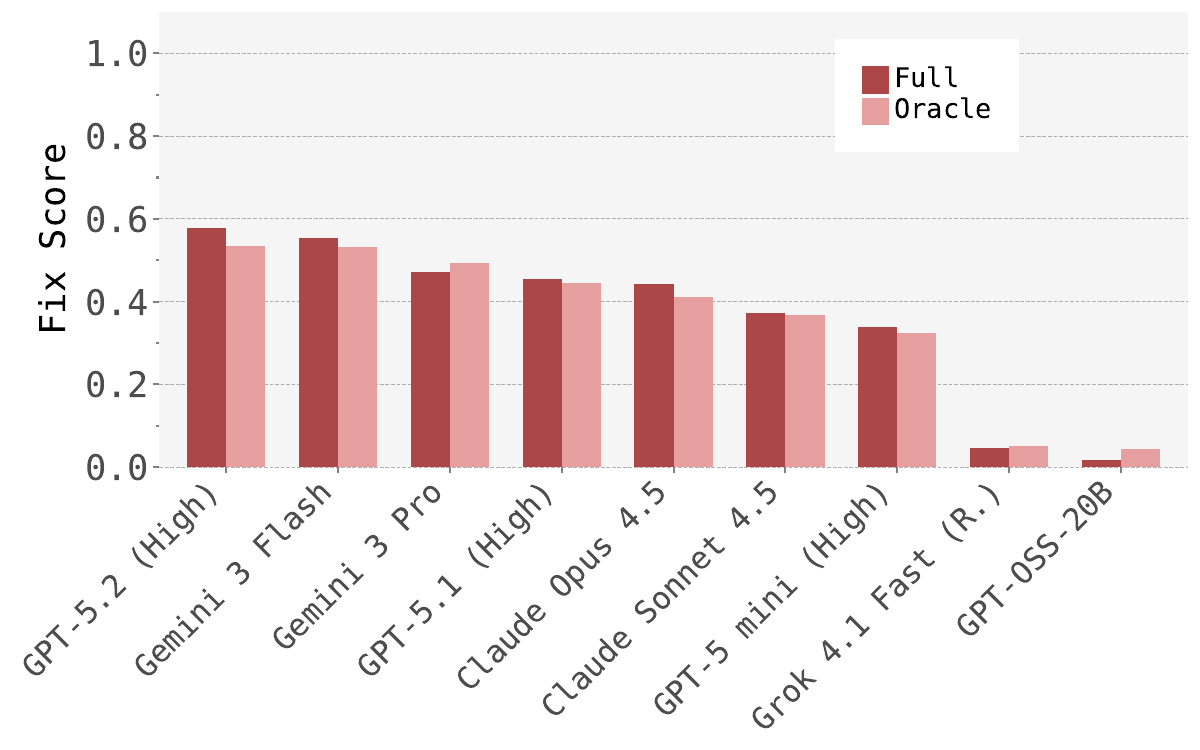}
    \caption{Frontier LLMs benefit from access to global context.}
    \label{fig:leaderboard_scores}
\end{figure}
% ------------ LEADERBOARD MAIN SCORE PLOT ------------

% ------------ SCORE DISTRIBUTION ------------
\begin{figure}[h]
    \centering
    \includegraphics[width=0.95\linewidth]{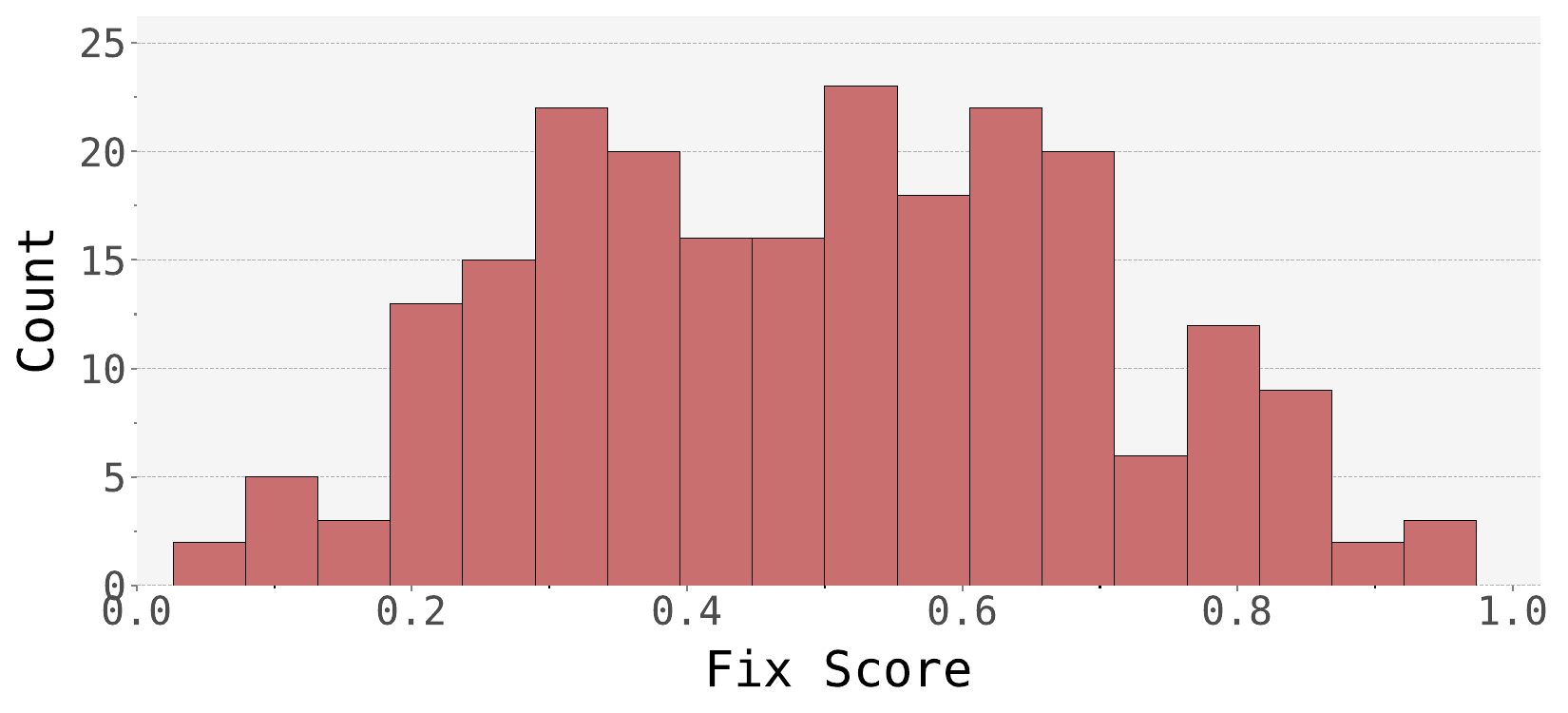}
    \caption{\system is not saturated with either impossible or trivial tasks.}
\label{fig:average_score_distribution}
\end{figure}
% ------------ SCORE DISTRIBUTION ------------

%\subsection{Main Results}
%\label{subsec:main_results}

Table~\ref{tab:main_leaderboard_table} and Figure~\ref{fig:leaderboard_scores} present the comprehensive evaluation of all 9 models on \system. The histogram of Fig.~\ref{fig:average_score_distribution} illustrates the distribution of fix scores \textemdash~calculated as the \textbf{per-task average} of the top five models. This quasi-normal distribution confirms a desirable property of the benchmark~\cite{hardt2025emerging}: it captures a spectrum of complexity rather than being saturated with impossible (score 0.0) or trivial (score 1.0) cases.

%Our analysis reveals the following trends regarding model capabilities.
%Although LLMs show potential in terms of problem diagnosis and repair metrics, they are able to fix misconfigurations to a full extent in less than 25\% of all cases only. This essentially indicates that LLM-centered systems still require continuous monitoring when scaled to challenging test environments. At the same time, they also prove useful as a powerful tool that can support real-world workflows given the human-in-the-loop condition.

While LLMs demonstrate potential for diagnostic and repair tasks, our analysis shows that they rarely produce fully correct fixes, with strictly correct resolutions (100\% fix rate with zero regressions) occurring in at most 25.5\% of cases. This ceiling in performance suggests that current models are best deployed as ``Human-in-the-Loop" assistants, consistent with recent research~\cite{holistic-incident-management} and industry practices~\cite{NetAssistant,BiAn,wang2025confucius}.

We detail the specific factors affecting LLM performance through the following insights:

\fakepar{Models perform better with access to global context} As shown in Fig.~\ref{fig:leaderboard_scores}, providing the \textit{full} network-wide configuration almost always outperforms the idealized \textit{oracle} setting, which only contains the faulty configuration files. This indicates that for top-performing models, the ability to examine configurations \textit{contextually} outweighs the noise introduced by all the irrelevant configuration data. We quantify this trade-off in the \textit{Retrieval mode} results (Table~\ref{tab:retrieval_mode}). When tasked with autonomous context selection, \textsc{GPT-5 mini} achieves high recall (82.6\%) of faulty routers, thereby effectively filtering noise and improving performance across all metrics. \textsc{Gemini 3.0 Flash}, however, misses critical configuration files (68.1\% Recall), and its performance degrades. This suggests that smaller models can benefit from \textit{careful} context selection. This finding is consistent with work on multi-agent systems powered by smaller language models~\cite{belcak2025smalllanguagemodelsfuture,wu2023autogenenablingnextgenllm}.

%This underscores the key challenge of identifying the \textit{relevant} context that points to the problem.

\newcommand{\dtext}[1]{\textbf{$\Delta\text{#1}$}}

%\subsection{Retrieval/(Context Caps)}
\begin{table}[!h]
\centering
\begin{tabular}{lccccc}
\toprule
\textbf{Model} & \textbf{Recall} & \dtext{Fix} & \dtext{Diag.}& \dtext{Regr.} & \textbf{$\Delta\$$}\\
\midrule
\hspace{-0.02cm}\geminiicon~\textsc{3 Flash} & 68.1\% & \textcolor{red}{-4.7\%} & \textcolor{red}{-2.0\%} & \textcolor{red}{+0.5\%} & +0.03\\
\gpticon~\textsc{5 Mini} & 82.6\% & \textcolor{ForestGreen}{+5.7\%} &  \textcolor{ForestGreen}{+4.0\%} & \textcolor{ForestGreen}{-2.0\%} & +0.01 \\
\bottomrule
\end{tabular}
\caption{Accurate retrieval of critical configurations can filter out noisy data and improve performance}
\label{tab:retrieval_mode}
\end{table}

\fakepar{Most efficacious LLMs are not always the safest} A high fix rate does not guarantee preservation of previously satisfied specifications. While \textsc{GPT-5.2} leads in fix rate (57.8\%), several other models achieve lower regression rates, including its predecessor \textsc{GPT-5.1}. In contrast, \textsc{Claude 4.5 Opus} is more conservative in its repairs, achieving a lower score of 44.2\% but the lowest regression rate at 3.6\%. 

\begin{figure}[h]
    \centering
    \includegraphics[width=1.01\linewidth]{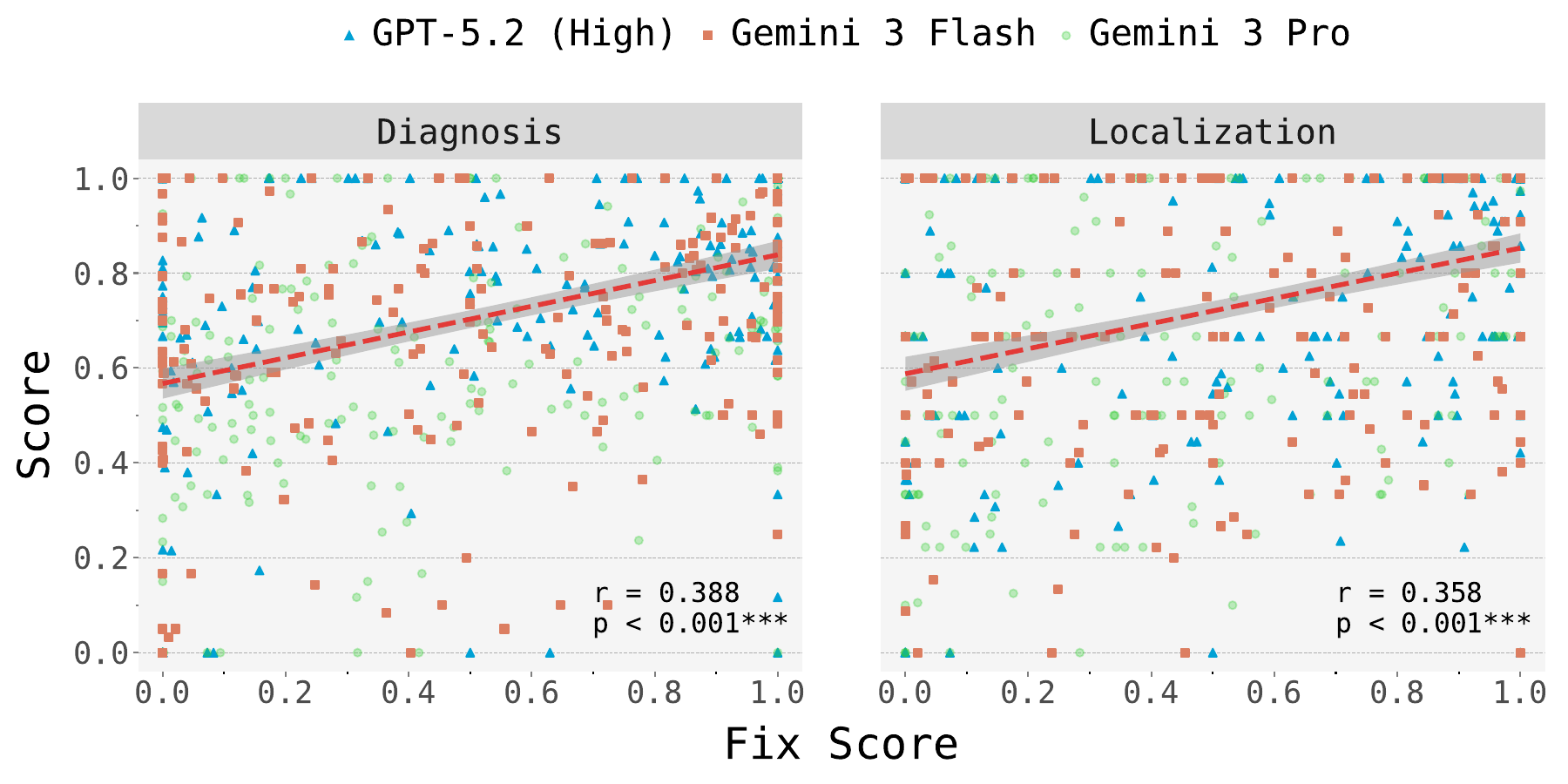}
    \caption{Diagnostic accuracy is necessary but not sufficient for repair.}
\label{fig:score_v_diag_corr}
\end{figure}

\fakepar{Accurate diagnoses lead to (but do not guarantee) effective fixes} We analyse the correlation between the diagnosis performance and final repair quality in Fig.~\ref{fig:score_v_diag_corr}. We observe a moderate positive correlation between diagnosis accuracy/localization and fix score. Crucially, the cluster of high diagnostic scores that yield poor repair metrics (upper-left quadrant in the scatter plot) represents cases in which models identified issues but failed to resolve them. Thus, while correct diagnoses often lead to correct fixes, they do not guarantee them.

% ------------ SCORE V DIAGNOSIS CORRELATION ------------

\begin{figure}[!h]
    \centering
    \includegraphics[width=1.01\linewidth]{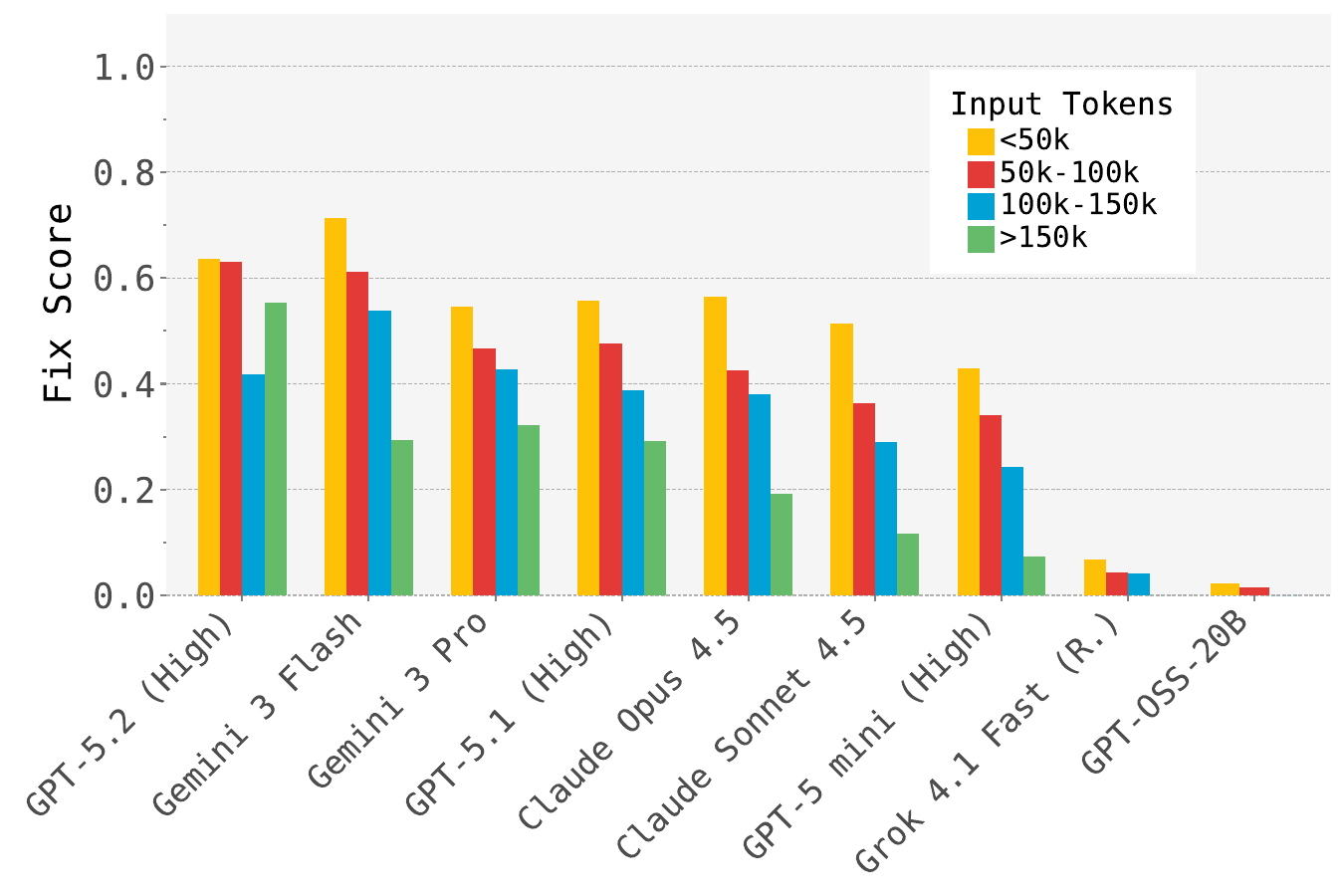}
        \caption{Repair performance consistently degrades with increasing context length.}
    \label{fig:impact_by_token_length}
\end{figure}

\fakepar{LLM performance degrades at scale} Topological scale, configuration length, and predicate set size are all factors that directly affect the amount of information included in the prompt. Hence, it is unsurprising that model performance consistently degrades with increasing input size, as shown in Fig.~\ref{fig:impact_by_token_length}. This amplifies the need for selecting relevant information for a model's limited context window.

\begin{figure}[h]
    \centering
    \includegraphics[width=1.01\linewidth]{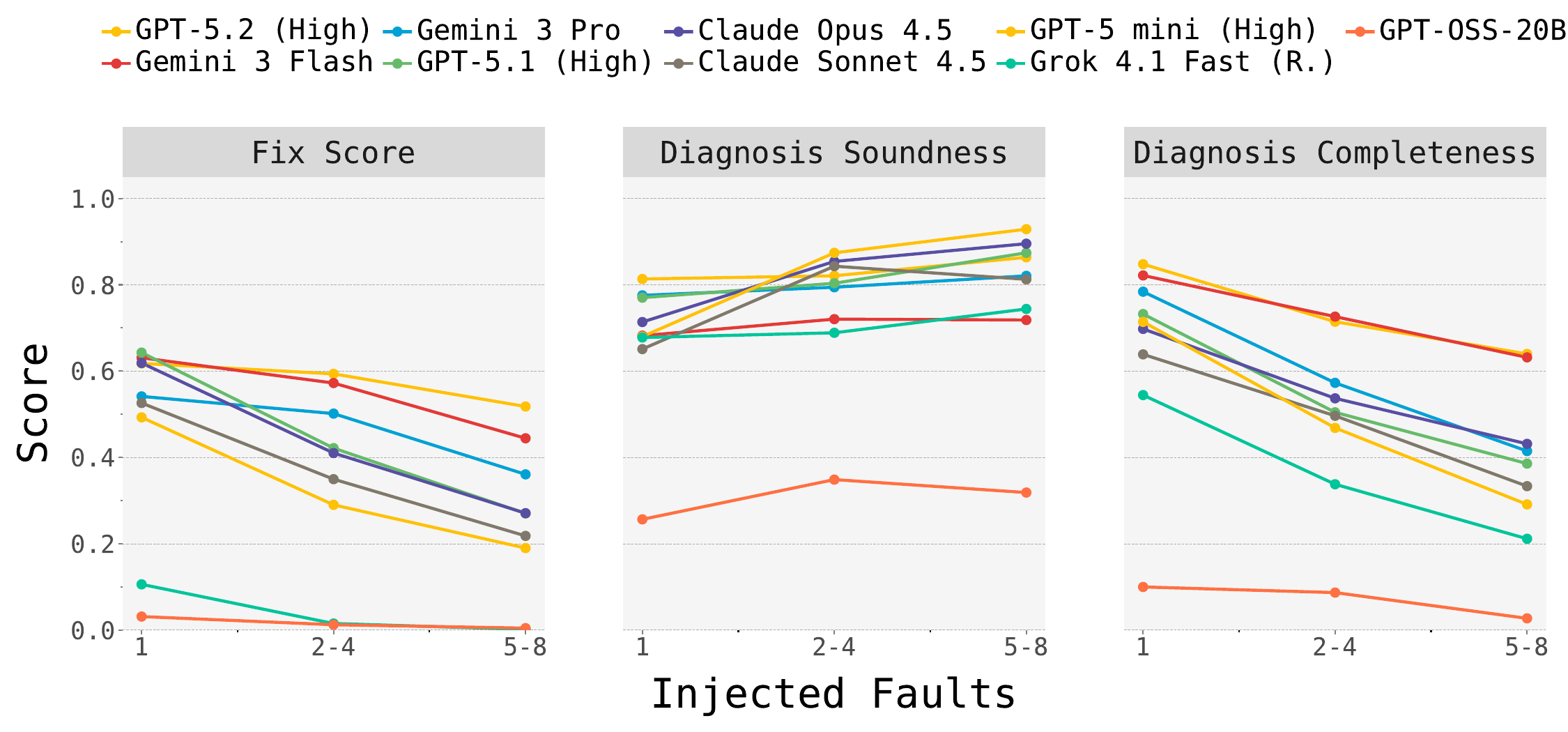}
        \caption{Models struggle to handle concurrent failures; as the number of root causes increases, diagnosis becomes partial and fix rate degrades.}
    \label{fig:impact_by_fault_count}
\end{figure}

\fakepar{LLMs struggle more to detect \textit{all} faults} We observe a revealing divergence between diagnostic \textit{soundness} (precision) and \textit{completeness} (recall) as fault multiplicity increases. As seen in Fig.~\ref{fig:impact_by_fault_count}, \textit{completeness} degrades sharply, while \textit{soundness} exhibits a slight upward trend. We hypothesize the following: In single-fault scenarios, the misconfiguration ``signal" is sparse, which often causes models to miss the issue entirely or even hallucinate faults \textemdash~hurting both soundness and completeness. Counterintuitively, as the number of faults increases, the abundance of such signals makes it easier for the model to identify \textit{any} of the real faults. Yet, models often stop at a partial diagnosis and reconfiguration, ignoring other disrupting misconfigurations.

\begin{figure}[ht]
    \centering
    \includegraphics[width=1.01\linewidth]{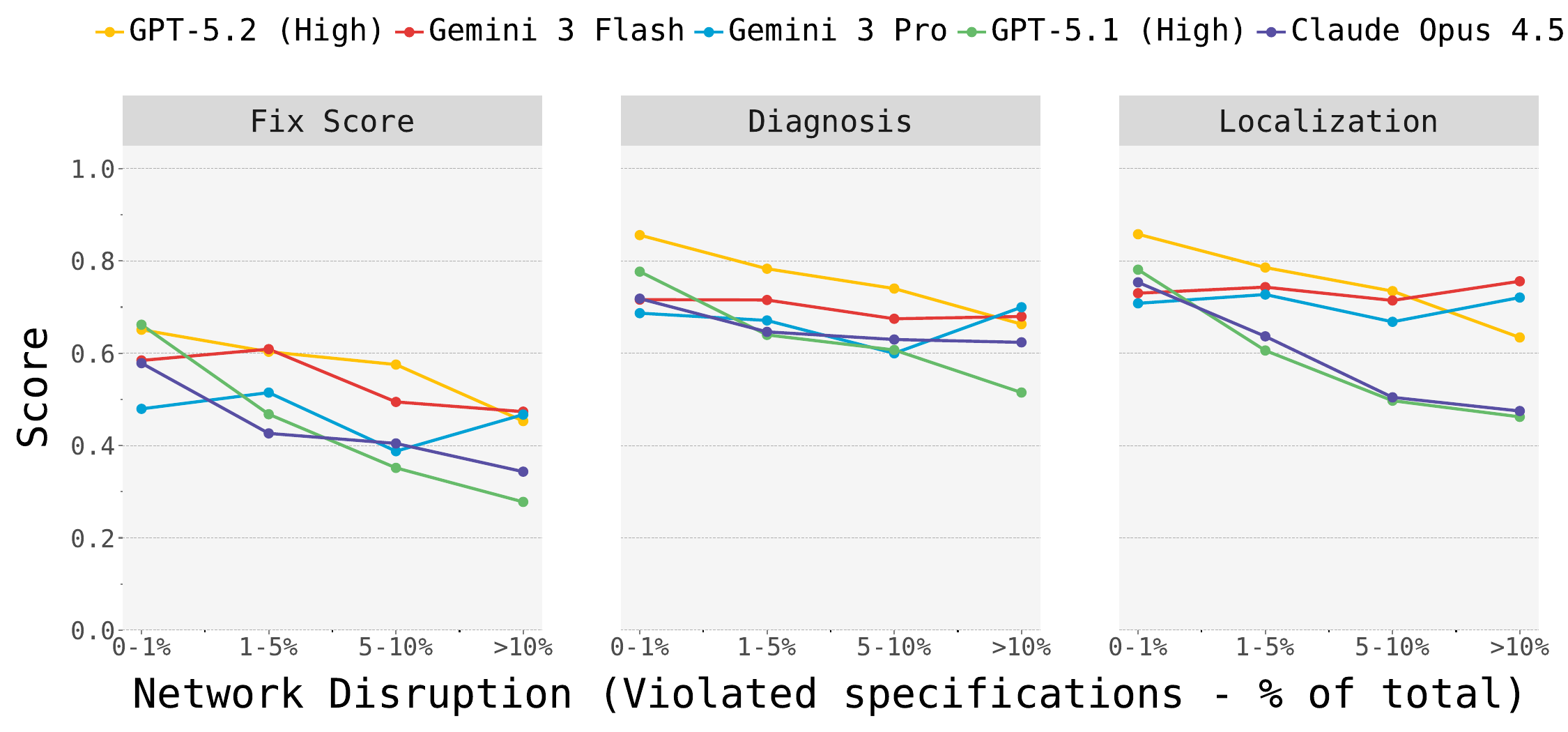}
        \caption{While some models remain robust, many perform poorly on more disruptive network faults.}
    \label{fig:impact_by_broken_predicates}
\end{figure}

\fakepar{Major network disruptions impact model performance} As shown in Fig.~{\ref{fig:impact_by_broken_predicates}}, we observe a general negative correlation between network disruption (percentage of broken predicates) and fix score performance. Notably, \textsc{GPT-5.1} exhibits the sharpest degradation, whereas more capable models such as \textsc{GPT-5.2} show resilience in repairing networks under more severe disruptions, indicating the ability to link broader specification violations to their root causes.

\fakepar{Takeaways} The results of our analysis show that:
\begin{itemize}[leftmargin=*]
    \item \textbf{Global context is critical but noisy}: While excessive data volume degraded performance, we found that models benefited from access to global configuration context. A system that effectively handles configuration repair should include a stage in which critical context is filtered in a dependency-aware manner, akin to recent work in~ \cite{jiang2024caipdetectingroutermisconfigurations}. 
    \item \textbf{Verification is a prerequisite for safety}: Regressions frequently accompany fixes, which is prohibitive while configuring networks. We posit that a system that automates configuration requires a closed loop with a verifier that proves the safety of solutions before deployment.
    \item \textbf{Iterative repair is needed for completeness}: Models struggle to resolve concurrent faults in a single pass, indicating that monolithic prompting fails at scale despite expansive context windows. Troubleshooting must be decomposed into an \textit{iterative agentic workflow} that diagnoses, proposes, and verifies solutions until the desired result is achieved.
\end{itemize}

%\fakepar{Input Tokens Length} To verify our hypothesis, we first illustrate the global effect of the increasing prompt tokens (see Fig.~\ref{fig:impact_by_token_length}). We note that performance scores consistently decrease almost for all models as the input prompts exhaust more tokens.

% ------------ SCALE IMPACT BY INPUT LENGTH ------------
% ------------ SCALE IMPACT BY INPUT LENGTH ------------

%\noindent \textit{Appendix B contains additional results.}

% ------------ FAULT IMPACT BY FAULT COUNT ------------
% ------------ FAULT IMPACT BY FAULT COUNT ------------

%\fakepar{Number of affected routers}

% % ------------ FAULT IMPACT BY ROUTER COUNT ------------
% \begin{figure}[ht]
%     \centering
%     \includegraphics[width=1.01\linewidth]{figures/fault_impact/fault_impact-None_mode-score_by_router_count.pdf}
%         \caption{Average main score change with respect to increasing number of affected routers in a task.}
%     \label{fig:impact_by_router_count}
% \end{figure}
% % ------------ FAULT IMPACT BY ROUTER COUNT ------------

% ------------ FAULT IMPACT BY BROKEN PREDICATES ------------
% ------------ FAULT IMPACT BY BROKEN PREDICATES ------------

% % ------------ CONTEXT FILLING ANALYSIS: RAND VS RAND+ORC + RETRIEVAL ------------
% \begin{figure}[ht]
%     \centering
%     \includegraphics[width=1.01\linewidth]{figures/context_filling/context_filling-random_caps_vs_retrieval-score.pdf}
%         \caption{.}
%     \label{fig:context_filling_analysis}
% \end{figure}
% % ------------ CONTEXT FILLING ANALYSIS: RAND VS RAND+ORC + RETRIEVAL ------------

% \subsection{}

\section{Related Work}
%\begin{itemize}
%\item Configuration synthesis and repair and verification.
%\item Agent and code benchmarks + Networking benchmarks
%\item Network troubleshooting
%\end{itemize}

\fakepar{Network verification} Two decades of research have established formal methods to mathematically prove network correctness. \textit{Data plane verification}~\cite{veriflow,anteater} checks that the network's forwarding behaviour satisfies some desired property, and \textit{control plane verification}~\cite{batfish,minesweeper}, verifies that a network configuration will produce a data plane that satisfies some intent~\cite{modelfree-krentsel}. \textit{Specification mining}~\cite{birkner2020config2spec} builds on these approaches to derive the set of forwarding specifications satisfied by a configuration. 

\system uses \textit{Batfish}~\cite{batfish} in conjunction with the specification mining algorithms proposed in \textit{Config2Spec}~\cite{birkner2020config2spec} as an integral component of its problem formulation. Specifically, we mine the ground-truth specifications from the correct reference network and use them to rigorously evaluate the functional correctness of the LLM-generated fixes.

%\fakepar{Automated configuration repair} 

\fakepar{LLM Benchmarks} Benchmarks such as SWE-Bench~\cite{swebench} and BaxBench~\cite{vero2025baxbenchllmsgeneratecorrect} indicate that rigorous evaluation of LLMs in software engineering faces fundamental challenges similar to those in network configuration repair. First, ensuring usability requires automatic, functional verification of solutions. Second, realistic tasks require navigating large volumes of noisy data (whether from entire code repositories or network-wide configurations) to identify useful information and arrive at a solution. \system contextualizes these challenges within the realm of network configuration repair, ensuring that it reflects the complexities of real-world network operations.

\fakepar{LLMs for network operations} Interest in LLMs to address the limitations of formal verification is growing, evidenced by recent industry assistants~\cite{NetAssistant, wang2025confucius,BiAn} and benchmarks~\cite{netconfeval,netllmbench,nika, netpress}. \system complements those efforts by rigorously evaluating end-to-end configuration repair, thereby advancing understanding of AI's potential and applicability to automated network operations. 
\section{Discussion and Limitations}

\newcommand{\question}[1]{\vspace{0.5em} \noindent \textit{\textbf{#1}} \xspace}

\question{What about evaluating fancier LLM-based systems?} The strategy space for LLM-based configuration repair is immense. Among others, it encompasses different \textit{prompting strategies} that affect a model's reasoning process~\cite{brown2020languagemodelsfewshotlearners, wei2023chainofthoughtpromptingelicitsreasoning}, \textit{retrieval methods} that determine which relevant information is included in the context from large volumes of data~\cite{lewis2021retrievalaugmentedgenerationknowledgeintensivenlp}, and \textit{agentic systems}, which constitute a distinct design space of their own~\cite{yao2023reactsynergizingreasoningacting,wu2023autogenenablingnextgenllm,hong2024metagptmetaprogrammingmultiagent}.

In this work, we assess the intrinsic capabilities of LLMs to reason about network state and resolve misconfigurations. We argue that establishing this baseline is critical, as it decouples performance attributable to the model's reasoning power from gains attributed to complex-system scaffolding. Still, we design \system as a modular platform that supports the evaluation of such advanced setups (including RAG and agentic systems).

\question{What about specifications under failures?} Network operators often care about invariants holding across multiple environments (e.g., maintaining reachability under any 2 link failures). While control-plane verification~\cite{minesweeper} can prove properties across all possible environments, we focus on specifications for a single environment; we posit that performance in this setting serves as a necessary upper bound on model capability. Since our results demonstrate that LLMs already struggle to reason about network state in a single environment, introducing the complexity of failure models is currently premature. Evaluating reasoning across all possible environments efficiently enough to support many benchmark cases will be considered in the future. 

\section{Conclusion}

We presented \system, a comprehensive framework for evaluating LLM-driven configuration repair. \system generates diverse scenarios and rigorously evaluates the end-to-end troubleshooting process by assessing diagnostic accuracy and formally verifying the correctness of repairs. Our evaluation of 9 state-of-the-art LLMs on 231 generated scenarios reveals their potential to diagnose misconfigurations and their struggle to reliably synthesize correct and safe reconfigurations. By providing a platform for thorough assessment of these repair capabilities, \system contributes towards the advancement of reliable, automated network operations.

\smallskip

\noindent \textit{This work does not raise ethical issues.}

\bibliographystyle{unsrt} 
\bibliography{references}

\clearpage
\appendix
\onecolumn
\section{Fault Library}
\label{appendix:fault-library}

\begin{center}
    
\centering
\scriptsize % Reduces font size slightly to fit the dense content; change to \small if space permits.
\captionof{table}{Comprehensive Fault Catalog listing the protocols affected, the nature of the misconfiguration (Summary), and the resulting impact on the network (Expected Effect).}
\label{tab:fault_catalog}
\begin{tabularx}{\textwidth}{l p{0.35\textwidth} X} % X column automatically fills remaining width
\toprule
\textbf{Protocol / Type} & \textbf{Summary} & \textbf{Expected Effect} \\
\midrule

% --- BGP Faults ---
\textbf{BGP} 
 & eBGP neighbor configured with incorrect remote AS 
 & eBGP session reset due to ASN mismatch, cutting off inter-AS route exchange \\
 
 & Administratively shut down a BGP neighbor 
 & BGP Peering is administratively disabled, withdrawing all prefixes learnt via the neighbor \\

& Node configured with incorrect local ASN 
 & Misaligned local ASN breaks iBGP/eBGP sessions and splits the AS control plane \\
 
 & Force invalid next-hop on eBGP advertisements 
 & Outbound policy rewrites next-hop to an unreachable address, causing downstream traffic blackholes \\
 
 & Remove next-hop-self from RR $\rightarrow$ client iBGP session 
 & iBGP routes advertised to clients retain original eBGP next-hop, which may be unreachable from clients causing traffic blackholes \\
 
 & Withdraw a BGP network statement from the process 
 & Prefix is no longer originated, withdrawing reachability from downstream peers \\
 
 & Remove outbound route-map from eBGP neighbor 
 & Export policy no longer enforced, allowing infrastructure routes (loopbacks, P2P) and unintended prefixes to leak to external peers \\
 
 & Swap inbound and outbound route-maps on a neighbor 
 & Inbound filters begin applying outbound and vice versa, breaking intended import/export policy \\
 
 & Leak router loopback by stripping export/import policies 
 & ASBR originates its loopback /32 into eBGP and the peer accepts it because inbound filtering was removed \\
 
 & Break RR sessions to orphan clients 
 & iBGP sessions removed between RR and up to 5 (exclusive) clients, orphaning them from iBGP reachability \\
 
 & Duplicate cluster-id across route reflectors and isolate clients on one RR 
 & Conflicting cluster-ids cause route reflectors to drop one another's updates, stranding clients that now depend on the misconfigured RR (cf. RFC 4456, Sec. 8) \\
\midrule

% --- OSPF Faults ---
\textbf{OSPF} 
 & OSPF interface cost set to extreme value 
 & Artificially high OSPF cost diverts traffic away from the link based on alternate SPF paths \\
 
 & Disable OSPF adjacency on a link 
 & Removing the link from OSPF prevents adjacency formation and withdraws LSAs learned across it \\
 
 & Node missing OSPF area membership 
 & Router withdraws from all OSPF areas, tearing down adjacencies and LSAs \\
 
 & Assign duplicate OSPF router-ID to multiple routers 
 & OSPF adjacencies fail or LSAs rejected due to router-ID collision, fragmenting OSPF domain and blackholing traffic \\
\midrule

% --- IS-IS Faults ---
\textbf{IS-IS} 
 & Disable IS-IS on an intra-AS link 
 & Removing the link from IS-IS prevents adjacency formation and withdraws LSPs learned across it \\
 
 & Demote a Level-1-2 IS-IS router to Level-1 
 & Reduces inter-area reachability by removing a backbone-capable router, risking L2 partitioning \\
 
 & Assign router to wrong IS-IS area 
 & Router in wrong area cannot form L1 adjacencies with its physical neighbors; causes partition of L1 domain and reachability loss \\
\midrule

% --- Basic/Addressing Faults ---
\textbf{Addressing} 
 & Duplicate loopback IPv4 addresses 
 & Two routers share the same loopback, risking routing loops and control-plane instability \\
 
 & Link interfaces disagree on prefix length 
 & One side of a point-to-point link uses a mismatched subnet mask, preventing adjacency formation \\
 
 & Link interfaces reside in different subnets 
 & Interfaces on a point-to-point link move to disjoint IPv4 subnets, breaking adjacency formation \\
\midrule

% --- Device Faults ---
\textbf{Device} 
 & Remove supporting static route for advertised prefix 
 & Advertised network disappears once the backing static route is withdrawn, causing a control-plane withdraw \\
\midrule

% --- Policy Faults ---
\textbf{Policy} 
 & Remove permit entry from prefix-list 
 & Prefix-list no longer matches intended prefixes, causing route filtering to block previously allowed routes \\
 
 & Convert BGP route-map permit clause into deny 
 & Previously exported prefixes are now filtered, withdrawing routes from neighbors \\
 
 & Lower BGP local-preference on inbound policy 
 & Reduced local-preference makes an alternate egress the best path for affected prefixes \\
\midrule

% --- Redistribution Faults ---
\textbf{Redistribution} 
 & Drop BGP $\rightarrow$ OSPF redistribution on an ASBR 
 & Internal OSPF loses external reachability because Type-5 LSAs are never originated \\
 
 %& Re-disable OSPF on inter-AS links 
 %& Dual-protocol adjacency collapses, preventing OSPF from running across AS boundaries \\
\midrule

% --- Security/ACL Faults ---
\textbf{Security} 
 & Insert implicit deny at top of interface ACL 
 & Ingress traffic on the protected interface is dropped before policy permits, breaking connectivity \\
 
 & Insert implicit deny at top of outbound interface ACL 
 & Egress traffic on the protected interface is dropped before policy permits, breaking connectivity \\

\bottomrule
\end{tabularx}
\end{center}

\clearpage
\section{Additional Results}
\label{appendix:more-plots}

% ------------ JUDGE, LOCALIZATION, REGRESSION PLOTS ------------
\begin{figure*}[!h]
    \centering
    % First subplot
    \begin{subfigure}[h]{0.33\textwidth}
        \centering
        \includegraphics[width=\textwidth]{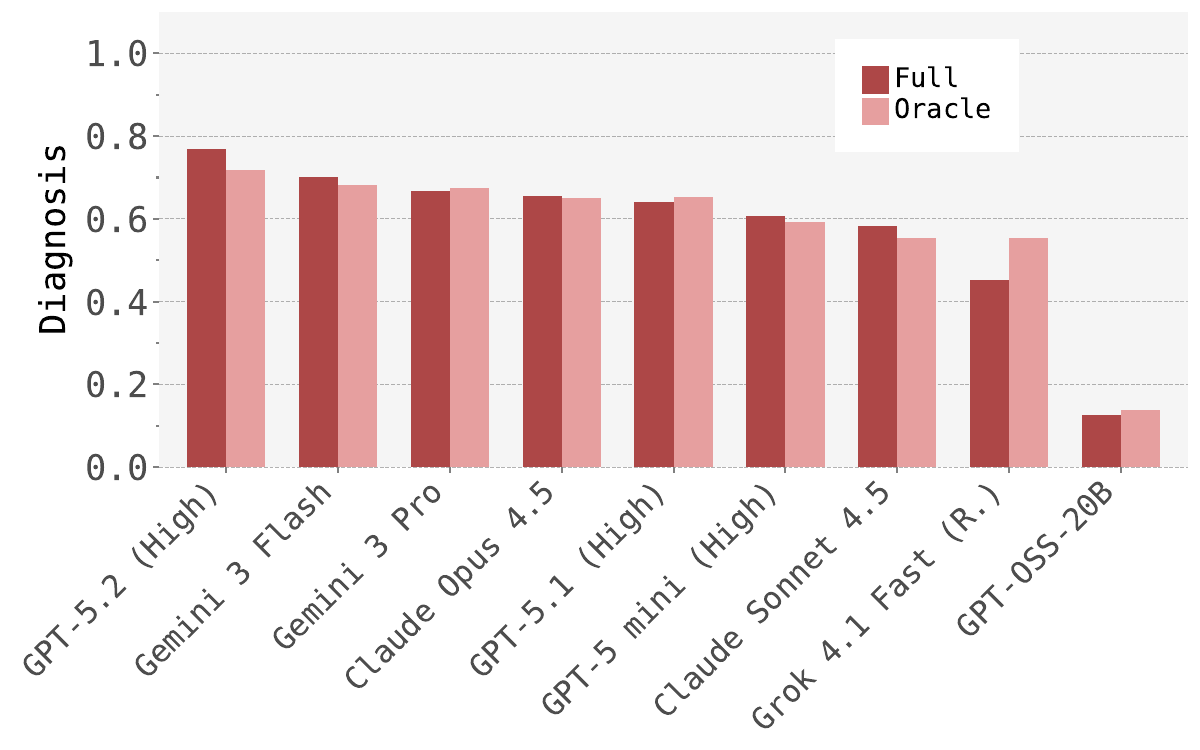} 
        \caption{Diagnosis}
    \end{subfigure}
    \hfill
    % Second subplot
    \begin{subfigure}[h]{0.33\textwidth}
        \centering
        \includegraphics[width=\textwidth]{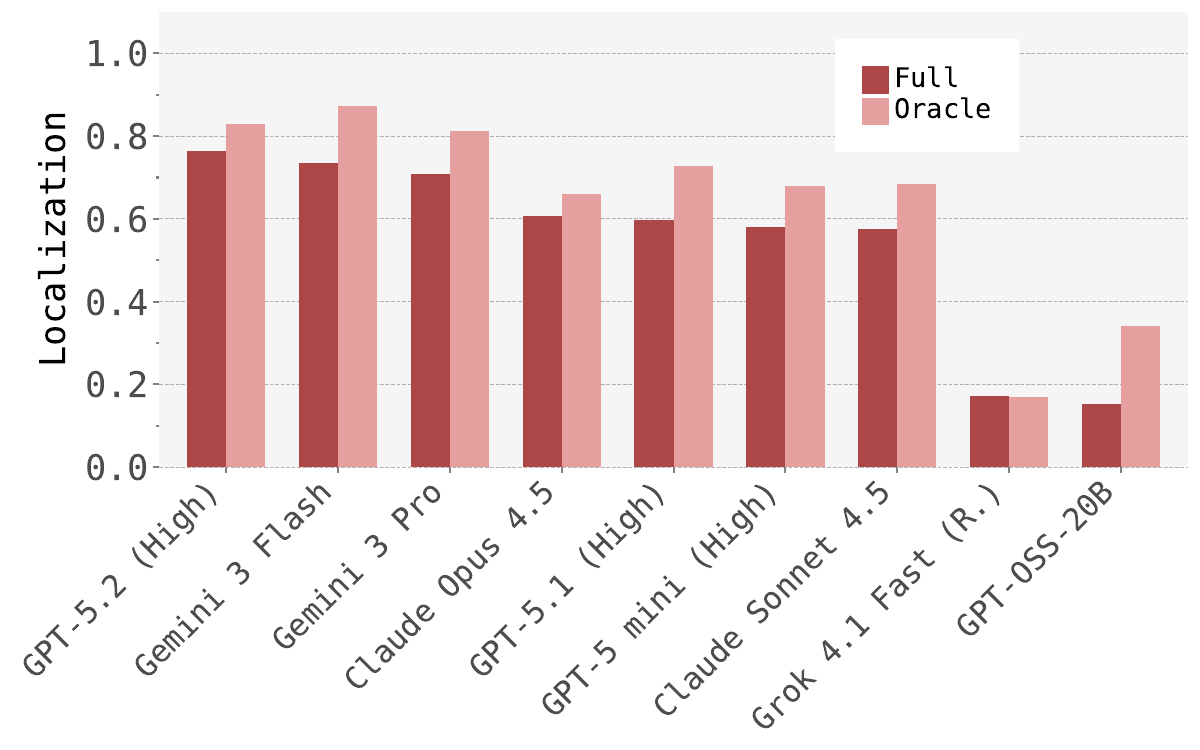} 
        \caption{Localization}
    \end{subfigure}
    \hfill
    % Third subplot
    \begin{subfigure}[h]{0.33\textwidth}
        \centering
        \includegraphics[width=\textwidth]{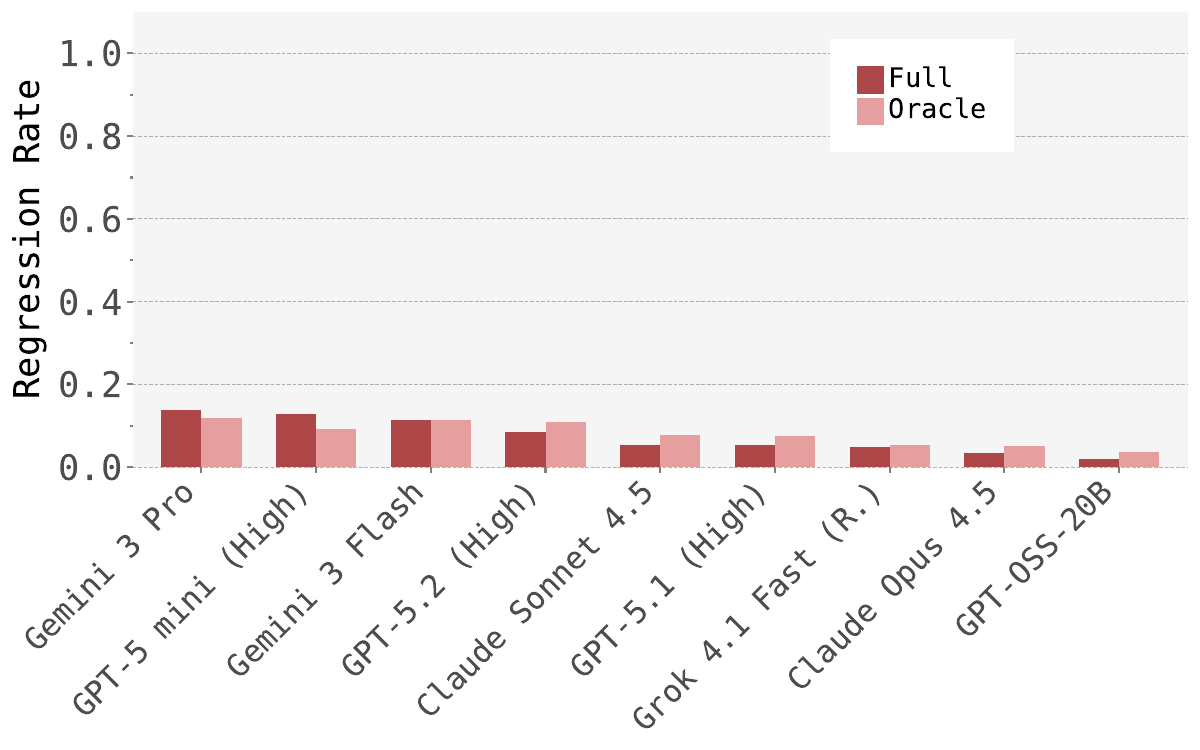} 
        \caption{Regression Rate}

    \end{subfigure}

    \caption{Overview of the model leaderboard using other core performance metrics: diagnosis (left) and localization (center) scores, followed by regression rate (right). }
    \label{fig:diag_local_reg_scores}
\end{figure*}
% ------------ JUDGE, LOCALIZATION, REGRESSION PLOTS ------------

% ------------ SCALE IMPACT BY TOKEN LENGTH (DIAGNOSIS, REG. RATE) ------------
\begin{figure*}[!h]
    \centering
    \begin{subfigure}[h]{0.43\textwidth}
        \centering
        \includegraphics[width=\textwidth]{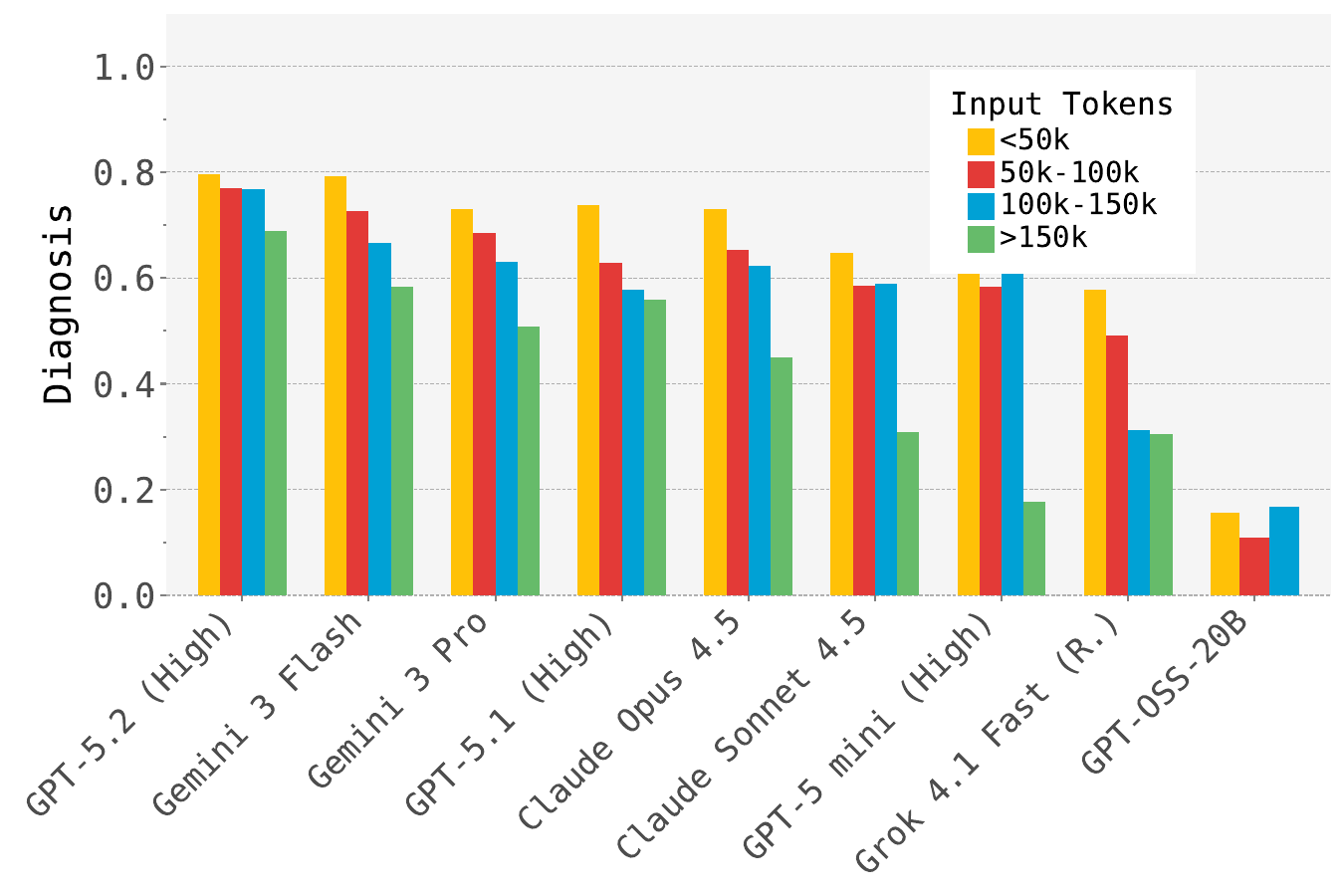}
        \caption{Diagnosis}
        \label{fig:scale_impact_token_v_diagnosis}
    \end{subfigure}%
    \hfill%
    \begin{subfigure}[h]{0.43\textwidth}
        \centering
        \includegraphics[width=\textwidth]{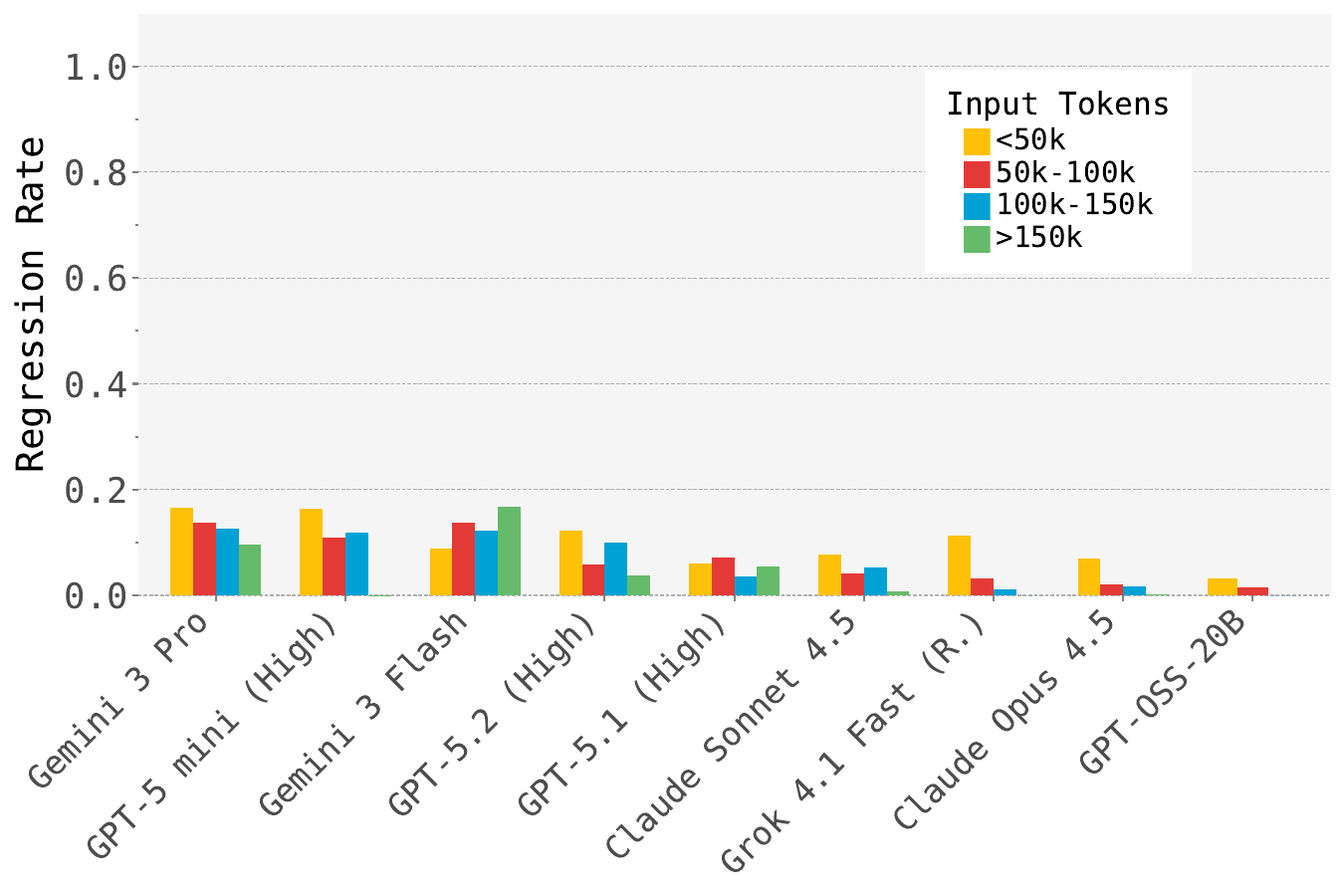}
        \caption{Regression Rate}
        \label{fig:scale_impact_token_v_regression_rate}
    \end{subfigure}%
    \caption{Diagnosis performance (left) consistently degrades with increasing input prompt tokens. The same trend is noticeable for regression rates (right) too; this potentially stems from the fact that with smaller context models might hallucinate more and break correct predicates.}
    \label{fig:pareto_cost_v_performance}
\end{figure*}
% ------------ SCALE IMPACT BY TOKEN LENGTH (DIAGNOSIS, REG. RATE) ------------

% ------------ PARETO FRONTIER PLOT (COST V PERFORMANCE) ------------
\begin{figure*}[!h]
    \centering
    \begin{subfigure}[h]{0.43\textwidth}
        \centering
        \includegraphics[width=\textwidth]{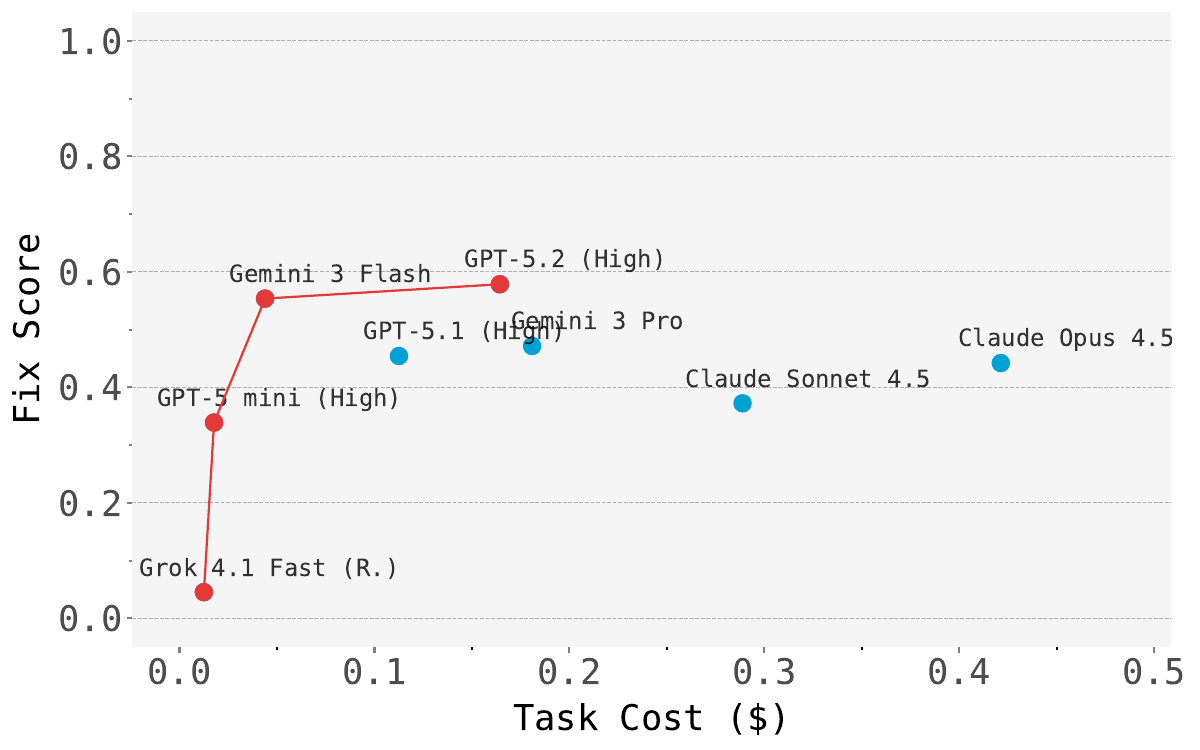}
        \caption{Fix Score}
        \label{fig:pareto_cost_v_main_score}
    \end{subfigure}%
    \hfill%
    \begin{subfigure}[h]{0.43\textwidth}
        \centering
        \includegraphics[width=\textwidth]{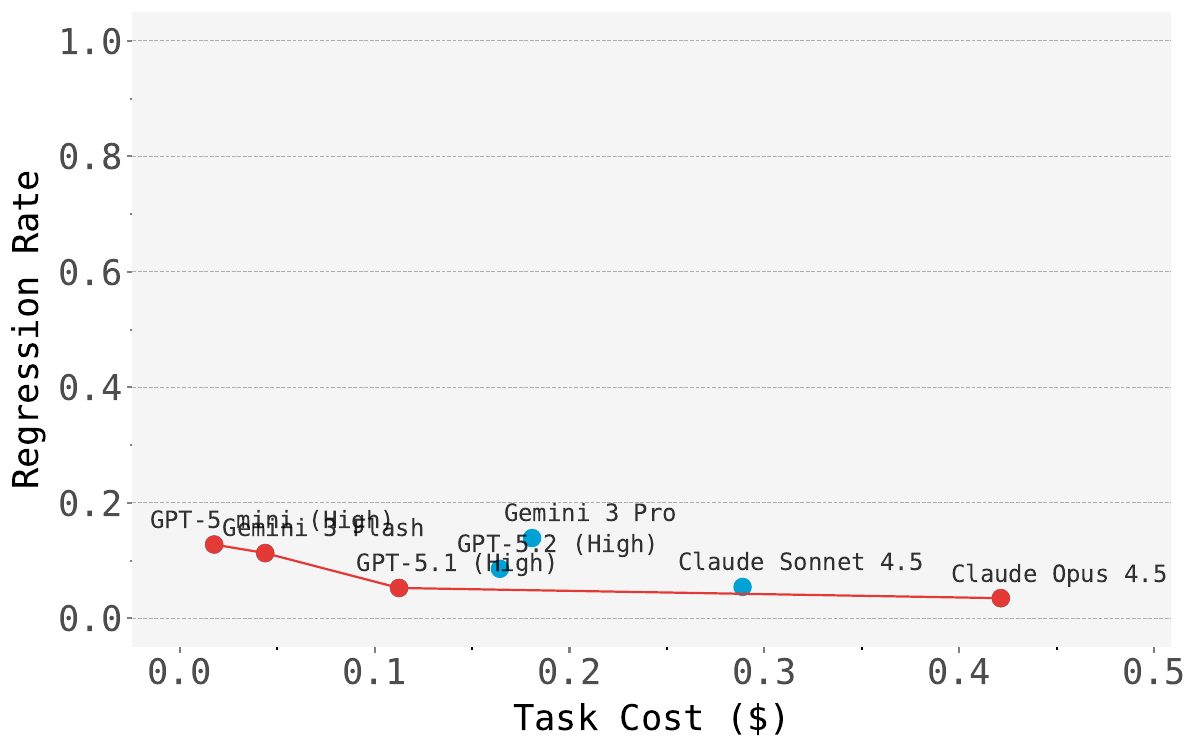}
        \caption{Regression Rate}
        \label{fig:pareto_cost_v_regression}
    \end{subfigure}
    \caption{Cost-Pareto frontier with respect to average fix score (left) and regression rate (right).}
    \label{fig:pareto_cost_v_performance}
\end{figure*}
% ------------ PARETO FRONTIER PLOT (COST V PERFORMANCE) ------------

\end{document}